\documentclass[journal]{new-aiaa}
\usepackage[utf8]{inputenc}
\usepackage{textcomp}
\usepackage{graphicx}
\usepackage{amsmath}
\usepackage[version=4]{mhchem}
\usepackage{siunitx}
\usepackage{longtable,tabularx}
\title{Reduction of Saturn Orbit Insertion Impulse using Deep-Space Low Thrust}

\author{Elena Fantino \footnote{Corresponding author}\footnote{Assistant Professor, Department of Aerospace Engineering, elena.fantino@ku.ac.ae.}}
\affil{Khalifa University of Science and Technology, Abu Dhabi, United Arab Emirates.}
\author{Roberto Flores\footnote{Associate Research Professor, Campus Norte UPC, Gran Capit\'an s/n, rflores@cimne.upc.edu.}}
\affil{International Center for Numerical Methods in Engineering, 08034 Barcelona, Spain.}
\author{Jes\'us Pel{\'a}ez\footnote{Full Professor, Space Dynamics Group, ETSIAE, Pza. Cardenal Cisneros 3, j.pelaez@upm.es.} and Virginia Raposo-Pulido\footnote{PhD, Space Dynamics Group, ETSIAE, Pza. Cardenal Cisneros 3, v.raposo.pulido@upm.es.}}
\affil{Technical University of Madrid, 28040 Madrid, Spain.}

\begin{document}

\maketitle

\begin{abstract}
Orbit insertion at Saturn requires a large impulsive manoeuver due to the velocity difference between the spacecraft and the planet. This paper presents a strategy to reduce dramatically the hyperbolic excess speed at Saturn by means of deep-space electric propulsion. The interplanetary trajectory includes a gravity assist at Jupiter, combined with low-thrust maneuvers. The thrust arc from Earth to Jupiter lowers the launch energy requirement, while an {\it ad hoc} steering law applied after the Jupiter flyby reduces the hyperbolic excess speed upon arrival at Saturn. This lowers the orbit insertion impulse to the point where capture is possible even with a gravity assist with Titan. The control-law algorithm, the benefits to the mass budget and the main technological aspects are presented and discussed. The simple steering law is compared with a trajectory optimizer to evaluate the quality of the results and possibilities for improvement.
\end{abstract}

\vspace{0.5cm}

\section{Introduction}
\lettrine{T}he giant planets have a special place in our quest for learning about the origins of our planetary system and our search for life, and robotic missions are essential tools for this scientific goal. Missions to the outer planets have been prioritized by NASA and ESA, and this has resulted in important space projects for the exploration of the Jupiter system (NASA's Europa Clipper \cite{phillips2014europa} and ESA's Jupiter Icy Moons Explorer \cite{grasset2013jupiter}), and studies are underway to launch a follow-up of Cassini/Huygens called Titan Saturn System Mission (TSSM)  \cite{2009TSSM}, a joint ESA-NASA project. 
Orbiter missions to Uranus and Neptune are also being considered \cite{2005SpilkerNeptune}: in this case, the amount of propellant required to decelerate and be captured by the planets' gravity on arrival is very large, and the support of techniques like aerobraking and aerocapture is being explored (see, e.g., Ref.~\cite{noca2004mission}). Cassini/Huygens travelled to Saturn following a $\Delta$V-VVEJGA\footnote{$\Delta$V stands for orbital maneuver, V for Venus, E for Earth, J for Jupiter, GA for gravity assist. So Cassini/Huygens executed two consecutive gravity assists with Venus, one with Earth, one with Jupiter and midcourse maneuvers.} trajectory. Orbit insertion (OI) was achieved with a bipropellant engine producing a velocity variation of 622 m/s at the cost of approximately 800 kg of propellant \cite{goodson1998cassini}. The subsequent pericenter raising maneuver consumed another 314 kg. Additionally, deep space maneuvers and course corrections before OI increased the mass budget by almost 1000 kg \cite{leeds1996cassini}. Clearly, the impact of these operations on the size and cost of the mission was considerable.
One alternative to reduce the cost of exploring the giant planets is to use low-energy transfers between libration point orbits \cite{Lo2001}. An application of this concept aided by low-thrust (LT) propulsion can be found in Ref.~\cite{Fantino2012}. While extremely energy-efficient, the very long duration of the transfer renders this approach impractical for realistic missions.
Another alternative is using the Lorentz-force interaction with a planetary magnetic field to produce thrust. In standard spacecraft (S/C), the interaction reduces to magnetic torques and only the effect on S/C attitude is relevant \cite{Fantino2003a,Fantino2003b}. However, by using an electrodynamic tether (ET), a significant thrust can be produced to assist in orbit insertion \cite{ET_Handbook2010}. Whereas use of the ET is readily possible for Jupiter \cite{sanmartin2016analysis}, the case for other outer planets presents issues because the efficiency of S/C capture with an ET is inversely proportional to the square of the magnitude of the planetary magnetic field, which is remarkably weak at Saturn, Uranus and Neptune. 

However, in the specific case of Saturn, if the relative velocity between the S/C and Saturn at the encounter is decreased sufficiently, the capture and final OI of the probe can still be achieved using the drag produced by the ET (see Ref.~\cite{sanmartin2018comparative}). The ET concept paves the way towards missions to explore Saturn and its moons with S/C masses below one tonne (the launch mass of Cassini/Huygens was 5600 kg). This enables use of smaller launchers, significantly reducing the overall mission cost. 
In the category of electric propulsion systems, electric thrusters are widely used nowadays to reduce propellant costs. LT was employed for the first time in interplanetary space by NASA's Deep Space 1 probe in 1998 \cite{Rayman1997DeepSpace1}. Since then, important missions to the inner solar system, such as NASA's Dawn \cite{Rayman2006Dawn}, JAXA's Hayabusa \cite{2015Hayabusa} and ESA-JAXA's BepiColombo \cite{NOVARA2002BepiColombo} have employed it to great success. 
The design and optimization of a LT interplanetary trajectory is a complex problem which has attracted interest since the early days of space exploration. The reader is referred to the fundamental work of \cite{1961Edelbaum,1991Lawden,2000Pollard} for the optimization of orbital maneuvers and direct transfers, whereas \cite{1968Bell,1998Betts,2010Conway} illustrate the theory and techniques of optimal control. When the trajectory includes multiple gravity assists, the optimization problem becomes much more complex (see, e.g., Ref.~\cite{2006Vasile}).

We present a method to reduce substantially the hyperbolic excess speed upon arrival at Saturn and, thus, the amount of fuel required for the OI manoeuver. The transfer from Earth to Saturn is a $\Delta V$-JGA trajectory, powered by electric propulsion. During the Earth-to-Jupiter transfer, an electric thruster imparts a constant tangential acceleration, whereas in the Jupiter-to-Saturn leg the guidance strategy forces the maximum instantaneous rate of decrease of the hyperbolic excess velocity upon arrival at Saturn. This steering law is reminiscent of the Q-guidance method for missile targeting, with the hyperbolic excess velocity taking the place of the velocity-to-be-gained vector \cite{bhat1987optimal,1982Battin}\footnote{Obviously, this is just a partial analogy because the excess speed is a scalar value.}. The practical implementation of the method is, in essence, an application of the classical gradient descent optimization technique \cite{Cauchy1847}: the thrust angle yielding the maximum rate of decrease of the  relative velocity is chosen at every point of the trajectory, resulting in a simple and effective control law.
The gravity assist with Jupiter is unpowered and modelled with the patched-conics method. Planetary orbits are assumed circular and coplanar, and the trajectory is also 2D, although the method can be easily extended to 3D. Outside of the spheres of influence of the planets, the trajectory is computed with the Sun-S/C two-body model, including the thrust of the propulsion system. The propulsion performance characteristics are those of the NASA Evolutionary Xenon Thruster (NEXT) \cite{HermanNEXT2010}. 

The originality of the work and the main goal of the paper lie in the use of deep-space electric propulsion to drastically decrease the cost of Saturn orbit insertion. This result paves the way to new opportunities to explore Saturn (and the outer planets in general), for example using small spacecraft or constellations of small probes capable of achieving capture with significantly reduced amounts of propellant, or taking advantage of alternative orbit insertion techniques such as those listed previously.
As a secondary topic, we explored the possibility of using the electric propulsion system to reduce the high cost of a direct Earth-to-Jupiter transfer.

The paper is organized as follows: we first present the design of the Earth-to-Jupiter portion of the trajectory, including the GA; then we illustrate the control law algorithm, and we apply it to the Jupiter-to-Saturn transfer. Next, we analyse the results, we make a comparison with a trajectory optimizer and we evaluate the technological feasibility (including the choice of the thruster and the generation of electrical power to operate it). Finally, the main conclusions are drawn. This paper builds on preliminary results presented at the $29^{th}$ AAS/AIAA Space Flight Mechanics Meeting \cite{fantino2019aas}. 

\section{Design of the low-thrust transfer from Earth to Jupiter}
\label{sec:EJ}
The Earth-to-Saturn trajectory designed in this work is of type $\Delta V$-JGA, which is an unusual choice. Schemes involving Earth and Venus GAs are much more common (see, e.g., \cite{1992DAmario,1998Petropoulos,2006Kemble,2013Ikeda}) because they allow to reduce launch $C_3$ (the square of the hyperbolic excess speed with respect to Earth, also known as characteristic energy). For example, the $C_3$ imparted to Cassini/Huygens to inject it toward Venus was 16.6 km$^2$/s$^2$ and the encounter with Jupiter occurred approximately three years after launch. We selected a direct transfer from Earth to Jupiter for its simplicity, but all the findings are applicable to a VVEJGA trajectory, substituting the deep-space impulsive maneuvers with LT arcs.

The minimum $C_3$ for a direct Earth-to-Jupiter transfer without any propulsion assistance is 77 (km/s)$^{2}$ (Ref.~\cite{C3book}). This corresponds to the first impulse of a Hohmann transfer between the orbits of the two planets. Such a large initial energy requires a powerful launcher and severely limits the mass injected. For example, a Delta IV Heavy (4050H) is only capable of accelerating a 1500 kg payload at this $C_3$ level \cite{NEXTSaturn}.

To improve the results of a previous study, which used a ballistic trajectory to Jupiter \cite{fantino2019aas}, we analyze the effect of continuously firing the electric thruster after launch. The goal is raising the aphelion of the initial ellipse until it reaches the orbit of Jupiter. An effective method is applying maximum thrust in the direction of motion. This yields the fastest increase of specific orbital energy and, therefore, of the semimajor axis. It is also possible to change the aphelion by increasing the eccentricity of the orbit, but this process is slower. This will be clearly illustrated by the Jupiter-to-Saturn transfer results presented later.

For trajectory analysis, we shall assume the electrical thruster imparts a constant acceleration of $2.5 \cdot 10^{-5}$ m/s$^2$, which is equivalent to a force of 25 mN acting on a mass of 1000 kg. In the case of the NEXT motor, this level of thrust corresponds to a specific impulse of 1400 s, a propellant flow rate of $1.85 \cdot 10^{-6}$ kg/s and an electrical input power of 600 W. In reality, at constant thrust, the acceleration would increase as propellant is used. However, to obtain a conservative estimate, our computations assume constant acceleration. This covers for possible propulsion system degradation as well as periods where electrical power is diverted from the motor to the communications system to establish a link with Earth. The NEXT thruster is designed for a minimum propellant throughput of 450 kg \cite{van2007lifetime}, with analysis indicating that actual capability is over 750 kg \cite{NEXTdeep}. At a flow rate of $1.85 \cdot 10^{-6}$ kg/s (57 kg per year), the minimum expected lifetime of the motor is close to 8 years. This, as we shall demonstrate, is well above the requirements for the complete Earth-Jupiter-Saturn transfer. Therefore, the operational life of the electric thruster is not a limiting factor.

To generate the Earth-to-Jupiter trajectories we built a table with combinations of launch $C_3$ and initial flight path angle $\gamma$ (the angle between Earth's orbital velocity and the S/C trajectory when it leaves the planet's sphere of influence). The flight path angle is considered positive when the radial component of the velocity is positive. 

\begin{figure}
	\centering\includegraphics[width=8cm]{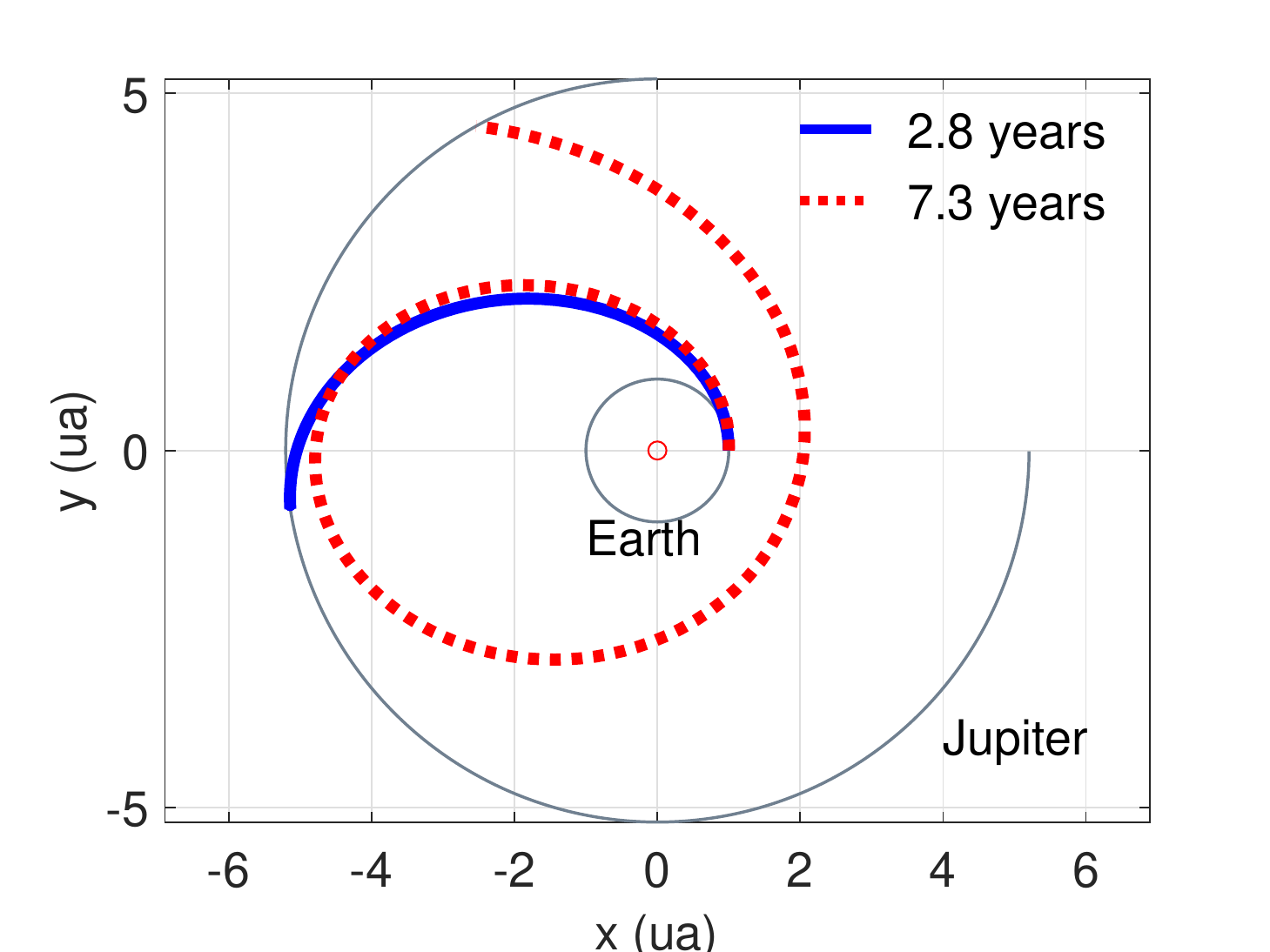}
	\caption{Short (solid blue) and long (dashed red) powered Earth-to-Jupiter transfers with launch $C_3$=67.25 (km/s)$^{2}$, $\gamma$=0$^{\circ}$ (solid blue) and $\gamma$=15$^{\circ}$ (dashed red), respectively.}
	\label{fig:EtoJ}
\end{figure}

We have explored values of $C_3$ between 65 and 72 (km/s)$^{2}$ in 0.25 (km/s)$^{2}$ steps. The range of $\gamma$ is from -15 to +15 degrees in 1 degree increments. Calculations show that, when $C_3$ is below 67.25 (km/s)$^{2}$, the transfer to Jupiter takes more than seven years. This happens because the S/C trajectory does not intersect Jupiter's orbit during the first passage through aphelion. Therefore, an additional revolution is needed to reach the planet, with a corresponding increase of transfer time. On the other hand, for $C_3$ values above the threshold, Jupiter can be reached in less than three years. Figure~\ref{fig:EtoJ} shows one long and one short Earth-to-Jupiter transfer. A seven years Earth-to-Jupiter transfer is impractical and would consume most of the operational life of the electric motor, so only those trajectories arriving at Jupiter in less than three years are retained for further study.

\begin{figure}
	\centering\includegraphics[width=8cm]{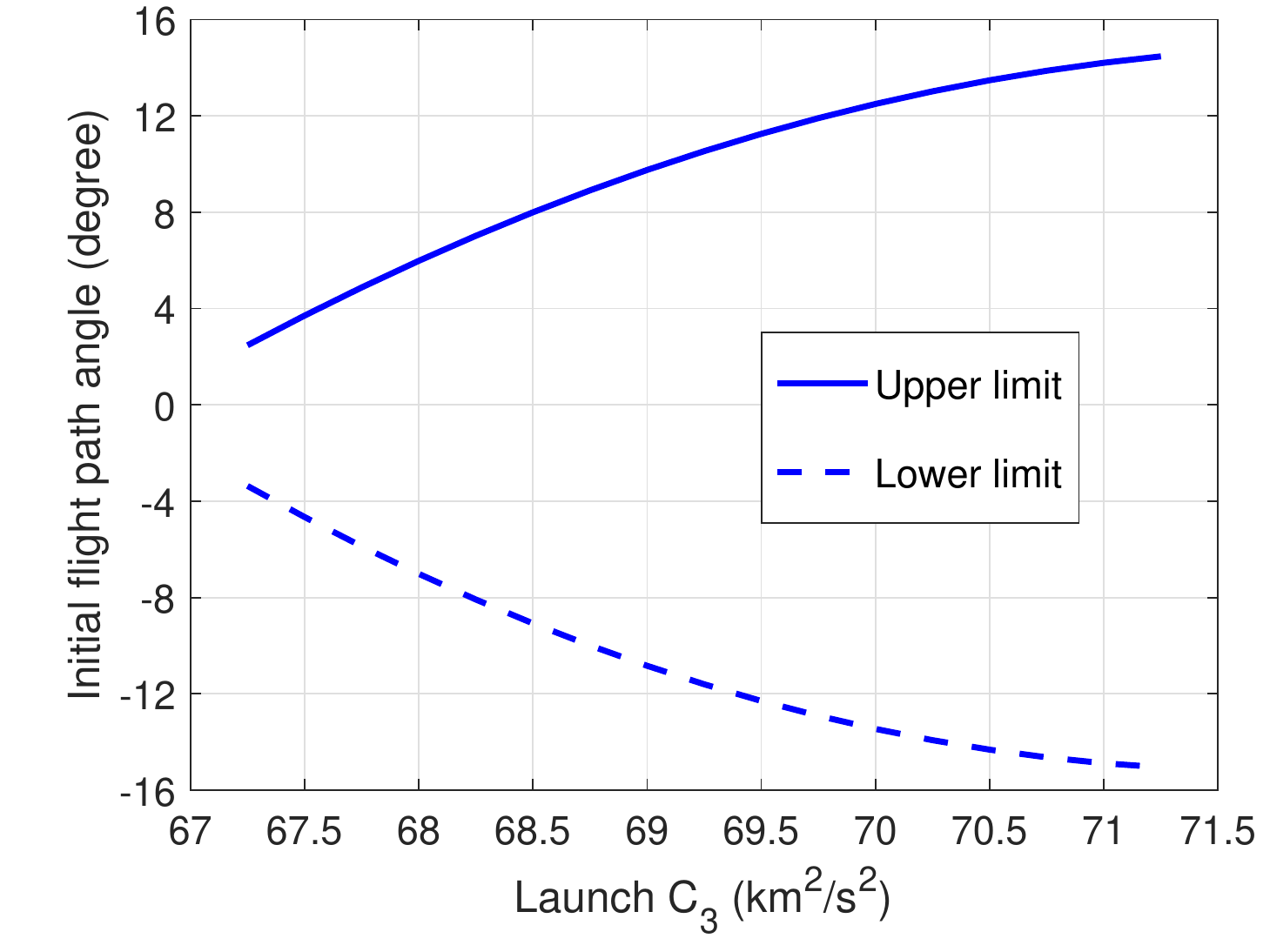}
	\caption{Maximum and minimum values of $\gamma$ compatible with a short powered Earth-to-Jupiter transfer.}
	\label{fig:fpa_c30}
\end{figure}

There is a range of values of the initial flight path angle for which the short transfer is possible. This range, centered on zero, becomes wider as $C_3$ increases. Figure~\ref{fig:fpa_c30} shows the bounds of $\gamma$ for which a short transfer is possible. Note that there is a slight asymmetry, as the magnitude of the lower bound is larger than the upper limit. It is due to trajectories with negative initial flight path angles passing closer to the Sun. This increases the orbital velocity for a given total energy and, consequently, the mechanical power of the thruster. Therefore, the rate of change of the semimajor axis is also increased making it possible to reach Jupiter in a shorter time. However, for moderate values of launch $C_3$ (i.e., close to the threshold) the range of $\gamma$ yielding a short transfer is quite narrow, so this effect is negligible in practice.

\begin{figure}
	\centering\includegraphics[width=8cm]{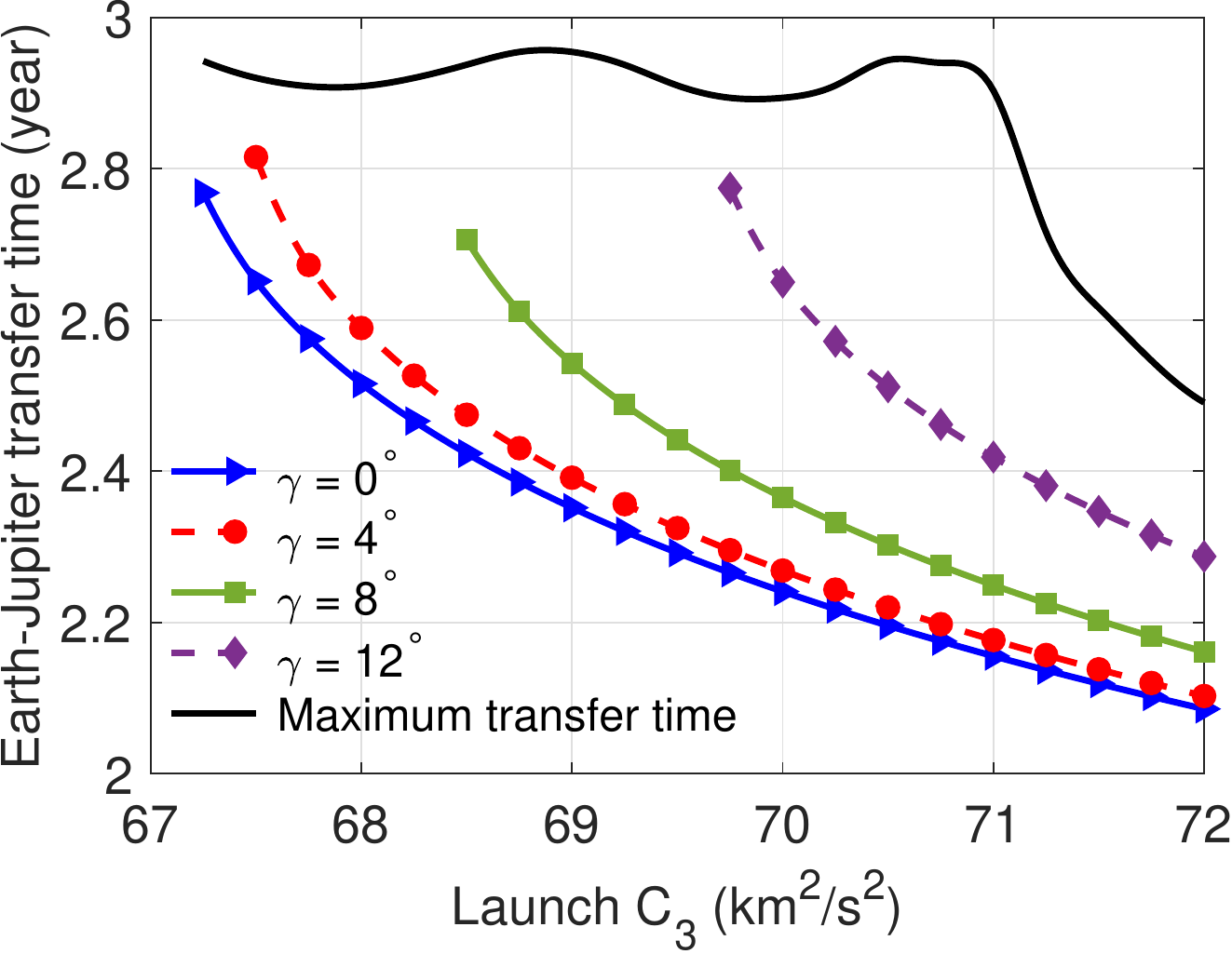}
	\caption{Earth-to-Jupiter transfer time vs. launch $C_3$ for short powered transfers.}
	\label{fig:t01_fpa_c30}
\end{figure}

Figure~\ref{fig:t01_fpa_c30} depicts the transfer times corresponding to the different combinations of flight path angle and launch energy.
For each value of $C_3$, the minimum time is obtained for $\gamma$=0 (i.e., departure tangent to Earth's orbit, maximizing initial mechanical energy). The minimum transfer time drops from 2.77 to 2.09 years if $C_3$ increases from 67.25 to 72 (km/s)$^{2}$. This reduces propellant consumption from 158 to 119 kg (a difference of 39 kg). However, assuming a payload mass sensitivity of 80 kg $\cdot$ s$^{2}$  /km$^{2}$ for the launcher (this quantity is the slope of the performance curve of Delta IV 4050H-19 for the $C_3$ range considered \cite{NEXTSaturn}), the larger $C_3$ reduces the injected mass by 380 kg. Therefore, for the Earth-to-Jupiter transfer, it is advantageous to use the minimum $C_3$, as the payload gain is an order of magnitude larger than the increase in propellant budget. The difference is so large that the same result holds for different launchers. Even for an Atlas V 551, which has a sensitivity 50\% smaller (\cite{NEXTSaturn}), the lowest $C_3$ remains the obvious choice. This is a natural consequence of the large specific impulse for the electric thruster. The maximum transfer time corresponds to the upper bound of the flight path angle range. The irregularities observed in the top curve of Figure~\ref{fig:t01_fpa_c30} are due to the discrete nature of the data set, combined with the high sensitivity of the transfer duration to the value of $\gamma$.

The Jupiter GA is modeled with the patched conics method. When the S/C crosses the orbit of Jupiter, the planet's velocity ${\bf V}_J$ is subtracted from the S/C velocity ${\bf V}_{S/C}^-$. The resulting entry hyperbolic excess velocity ${\bf v}_{\infty}^-$ and the choice of perijove radius $r_{\pi}$ fully determine the Jupiter-centered hyperbola, in particular the relative velocity ${\bf v}_{\infty}^+$ at sphere of influence exit and the corresponding new heliocentric velocity ${\bf V}_{S/C}^+$ of the S/C. We assume that both the velocity of Jupiter and the heliocentric position of the S/C remain unchanged during the GA. For each arrival trajectory we tested a range of flyby depths from 0.5 to 9.5 million kilometers in steps of 500 000 km. The set of post-GA states is used to provide the initial conditions for the Jupiter-to-Saturn trajectory design strategy discussed in the next section.

\section{A simple control law for low-thrust transfers}
\label{sec:LT}
\subsection{Nomenclature}
In a Cartesian inertial reference system, let us define the state vector ${\bf s}$ of a point-particle as
\begin{equation}
{\bf s }(t) = \left[ {\begin{array}{*{20}{c}}
{\bf r}\\
\dot{\bf r}\\
\end{array}} \right] = \left[ {\begin{array}{*{20}{c}}
x\\
y\\
{{v_x}}\\
{{v_y}}
\end{array}} \right]  .
\label{eq:1.1}
\end{equation}
When appropriate, we shall use polar coordinates $\left\{ {r,\theta } \right\}$  such that
\begin{equation}
{\bf{r}} = r{{\bf{u}}_r}\,\,\,;\,\,\,\,{{\bf{u}}_r} = \left[ {\begin{array}{*{20}{c}}
{\cos \theta }\\
{\sin \theta }
\end{array}} \right]\,\,\,;\,\,\,\,{{\bf{u}}_\theta } = \left[ {\begin{array}{*{20}{c}}
{ - \sin \theta }\\
{\cos \theta }
\end{array}} \right].
\label{eq:12}
\end{equation}
The particle moves under the gravitational attraction of a central body, characterized by its gravitational parameter $\mu$. The central body is located at the origin of the reference frame. Furthermore, there is a propulsion system capable of producing a specific thrust (i.e., acceleration) ${\bf{f}}$: 
\begin{equation}
{\bf{f}}(t) = {f_r}{{\bf{u}}_r} + {f_\theta }{{\bf{u}}_\theta }.
\label{eq:1.3}
\end{equation}
Thus, the system of governing equations is
\begin{equation}
\frac{{{\rm{d}}{\bf{s }}}}{{{\rm{d}}t}} = \left[ {\begin{array}{*{20}{c}}
{{v_x}}\\
{{v_y}}\\
{{a_x}}\\
{{a_y}}
\end{array}} \right]{\rm{   where}}\,\,{\bf{a}}{\rm{ = }}\left( {{f_r} - \frac{\mu }{{{r^2}}}} \right){{\bf{u}}_r} + {f_\theta }{{\bf{u}}_\theta }.
\label{eq:1.4}
\end{equation}
At each point in time, we shall define a vector of osculating Keplerian elements
\begin{equation}
{\bf o}({\bf s}) = \left[ {\begin{array}{*{20}{c}}
a\\
e\\
\omega \\
{{M_0}}
\end{array}} \right].
\label{eq:1.5}
\end{equation}
The components of the vector are the semimajor axis $a$, the eccentricity $e$, the longitude of the pericenter $\omega$ and the mean anomaly $M_0$ at epoch, respectively. In the following, we shall focus on $a$ and $e$ exclusively, but the procedure is applicable to all four elements.

\subsection{Problem statement}
Given a system governed by Eq.~(\ref{eq:1.4}), an initial state vector ${\bf s}(0) = {\bf s}_0$ and an error function $\Im$ expressed in terms of the osculating orbital elements (i.e., $\Im ({\bf{o}})$); find a control law ${\bf f}(t)$ that achieves the maximum reduction of the error function in a given time, subject to the constraint $\mid\mid {\bf f} \mid\mid \le f_{max}$  (i.e., the available thrust is fixed).

\subsection{Control law} 
We shall assume that maximizing the instantaneous rate of reduction of an error function  $\Im (a,e)$ is an effective guidance strategy. 
What follows is a direct application of the classical gradient descent optimization technique \cite{Cauchy1847}. In the context of modern guidance algorithms, it can be considered a particular case of the general Proximity Quotient guidance law (Q-Law) \cite{2004Petropoulos, 2005SeungwonB, 2014Falck}.

The thrust $f_{max}$ is an order of magnitude smaller than the gravitational pull of the Sun. Therefore, the effect of the propulsion system can be linearized (i.e., the rate of change of $\Im$ becomes a homogenous function of the thrust components). Thus, to achieve the fastest reduction of $\Im$, the maximum thrust has to be used at all times. The only free parameter is the orientation of the thrust vector ($\beta$)
\begin{equation}
{\bf{f}} = {f_{\max }}\left( {\sin \beta \,{{\bf{u}}_r} + \cos \beta \,{{\bf{u}}_\theta }} \right).
\label{eq:3.1}
\end{equation}
The control law is determined by the conditions
\begin{equation}
\frac{\partial }{{\partial \beta }}\left[ {\frac{{{\rm{d}}\Im \left( {{\bf{o}}({\bf{s }})} \right)}}{{{\rm{d}}t}}} \right] = 0\,\,\,\,{\rm{and}}\,\,\,\,\frac{{{\rm{d}}\Im \left( {{\bf{o }}({\bf{s }})} \right)}}{{{\rm{d}}t}} < 0.
\label{eq:3.2}
\end{equation}
We compute the rate of change of the error using the chain rule:
\begin{equation}
\frac{{{\rm{d}}\Im \left( {{\bf{o}}({\bf{s }})} \right)}}{{{\rm{d}}t}} = \nabla \Im ^T \cdot \frac{{{\rm{d}}{\bf{o }}}}{{{\rm{d}}t}}.
\label{eq:3.3}
\end{equation}
For our particular case, where  $\Im ({\bf{o }}) = \Im (a,e)$, the gradient is
\begin{equation}
\nabla \Im  = \left[ {\begin{array}{*{20}{c}}
{\frac{{\partial \Im }}{{\partial a}}}\\
{\frac{{\partial \Im }}{{\partial e}}}
\end{array}} \right].
\label{eq:3.4}
\end{equation}
The rate of change of the orbital elements can be expressed in matrix form as
\begin{equation} 
\frac{{{\rm{d}}{\bf{o }}}}{{{\rm{d}}t}} = \left[ {\begin{array}{*{20}{c}}
{\dot a}\\
{\dot e}
\end{array}} \right] = \displaystyle \left[ {\begin{array}{*{20}{c}}
{\dfrac{{\partial \dot a}}{{\partial {f_r}}}}&{\dfrac{{\partial \dot a}}{{\partial {f_\theta }}}}\\
{\dfrac{{\partial \dot e}}{{\partial {f_r}}}}&{\dfrac{{\partial \dot e}}{{\partial {f_\theta }}}}
\end{array}} \right] \cdot \left[ {\begin{array}{*{20}{c}}
{{f_r}}\\
{{f_\theta }}
\end{array}} \right] = {\bf{R}} \cdot {\bf{f}}.
\label{eq:3.5}
\end{equation}
The components of matrix ${\bf R}$ are computed from Gauss' planetary equations \cite{1812Gauss}
\begin{eqnarray}
\frac{{\partial \dot a}}{{\partial {f_r}}} & = & {c_1}\,e\,\sin \nu \,,\\
\frac{{\partial \dot a}}{{\partial {f_\theta }}} & = & {c_1}\left( {1 + e\cos \nu } \right)\,,\\
\frac{{\partial \dot e}}{{\partial {f_r}}} & = & {c_2}\sin \nu \,,\\
\frac{{\partial \dot e}}{{\partial {f_\theta }}} & = & {c_2}\left( {\cos \nu  + \cos E} \right)\,
\label{eq:3.6}
\end{eqnarray}
where
\begin{equation}
{c_1} = \frac{{2ah}}{{\mu \left( {1 - {e^2}} \right)}}\,\,{\rm;}\,\,{c_2} = \frac{h}{\mu }.\label{eq:3.7}
\end{equation}
In the expressions above, $\nu$ is the true anomaly, $E$ is the eccentric anomaly and $h$ denotes the specific orbital angular momentum.
Taking the derivative of Eq.~(\ref{eq:3.1}) with respect to $\beta$ gives
\begin{equation}
\frac{{\partial {\bf{f}}}}{{\partial \beta }} = {f_{\max }}\left[ {\begin{array}{*{20}{c}}
{\cos \beta }\\
{ - \sin \beta }
\end{array}} \right].
\label{eq:3.8}
\end{equation}
Combining Eqs.~(\ref{eq:3.2}), (\ref{eq:3.3}), (\ref{eq:3.5}) and (\ref{eq:3.8}) yields the relationship for the optimal thrust angle:
\begin{equation}
\nabla \Im^T \cdot {\bf{R}} \cdot \frac{{\partial {\bf{f}}}}{{\partial \beta }} = 0.\label{eq:3.9}
\end{equation}
Let  $\nabla \Im^T \cdot {\bf{R}} = \left[ {\begin{array}{*{20}{c}}
{{b_1}}&{{b_2}}
\end{array}} \right] = {{\bf{b}}^T}$. Vector ${\bf b}$ is a function of the current state of the system and does not depend on the thrust setting. Its components can thus be directly computed from the instantaneous state vector. Once ${\bf b}$ is known, Eq.~(\ref{eq:3.9}) can be solved for the thrust angle
\begin{equation}
{b_1}\cos \beta  - {b_2}\sin \beta  = 0\,\,\, \to \,\,\,\beta  = \arctan \frac{{{b_1}}}{{{b_2}}}.
\label{eq:3.10}
\end{equation}
Equation~(\ref{eq:3.10}) gives two distinct values of $\beta$ spread $180^{\circ}$ apart. The correct one is determined from the condition
\begin{equation}
{\bf{b}}^T \cdot {\bf{f}} < 0\,\,;
\end{equation}
the other value obviously corresponds to the maximum rate of increase of the error.

\section{Low-thrust transfer from Jupiter to Saturn: excess speed reduction strategy}
\label{sec:strat} 
The orbital periods of Jupiter and Saturn are 11.9 and 29.5 years, respectively. Therefore, the synodic period of the two planets is 19.9 years. For a given Jupiter-to-Saturn transfer, a long time may be required for the planets to reach a suitable configuration (i.e., allowing the S/C to actually encounter Saturn). Therefore, in a real mission, the departure window must be considered carefully when designing the trajectory.
In this preliminary analysis, however, the objective is to analyze the effectiveness of the guidance strategy in reducing the hyperbolic excess velocity on arrival at Saturn. Therefore, we assumed that the error function depends only on the semimajor axis and eccentricity of the heliocentric orbit, i.e. $\Im (a,e)$, and no phasing requirements are in place. For the sake of completeness, however, the results section contains summary information on the duration of the launch window.  

Upon exiting the sphere of influence of Jupiter, the electric thruster is powered continuously, delivering the same constant acceleration of $2.5 \cdot 10^{-5}$ m/s$^2$ as in the first portion of the interplanetary transfer. The thrust direction is continually adjusted using the strategy described in the previous section. The hyperbolic excess velocity upon Saturn arrival is used as the error function, causing the algorithm to reduce the relative speed of the S/C when it intersects the planet's orbit. Once this velocity falls below a predefined threshold, the motor is switched off and the S/C coasts to Saturn. We have limited the maximum duration of the post-Jupiter thrust arc to 4 years. This keeps the total time of operation of the propulsion system under 7 years, which is within its design limits.

It would be possible to apply the thrust arc anywhere between Jupiter and Saturn, or even use multiple shorter powered segments. However, it is advisable to use the electric motor as soon into the mission as possible for two operational reasons. First and foremost, as discussed later, the only viable electric generators are radioisotope thermoelectric generators (RTGs). These degrade with time, even in periods when no power is used. Given the long duration of the mission (over 12 years, as we shall see) it is advisable to power the motor as early as possible. Second, maximum reliability is achieved if the propulsion system works continuously at constant regime. Therefore, it is desirable to keep the motor working all the time until the final cutoff.

For the sake of simplicity, we shall use the square of the excess velocity as error function
\begin{equation}
\Im (a,e) = V_{ex}^2 = \left\| {{\bf{v}} - {{\bf{v}}_S}} \right\|_{r = {r_S}}^2,
\end{equation}
where the subindex $S$ denotes values of Saturn's orbit:
\begin{equation}
{r_S} = 9.537 \,{\rm au}\,\,\,;\,\,\,{{\bf{v}}_S} = \sqrt {\frac{\mu }{{{r_S}}}} {{\bf{u}}_{\theta}}.
\end{equation}
To improve the accuracy of the calculations, it is advisable to use normalized variables. In our case, we use as reference length the astronomical unit and as reference time one year. The dimensionless value of the gravitational parameter of the Sun is thus $4{\pi ^2}$. The square of the velocity of the probe upon intersecting Saturn's orbit is
\begin{equation}
{V^2} = \mu \left( {\frac{2}{{{r_S}}} - \frac{1}{a}} \right).
\label{eq:5.5}
\end{equation}
Note that Eq.~(\ref{eq:5.5}) is only physically meaningful if the trajectories of the probe and planet actually intersect, which is not a given. However, for the time being, we shall assume that this is the case. The situation when there is no intersection will be addressed later with a minor change of the expressions. From the conservation of angular momentum, the circumferential velocity of the probe at intersection is
\begin{equation}
V_{\theta} ^2 = \frac{{\mu a\left( {1 - {e^2}} \right)}}{{r_S^2}}.
\label{eq:5.6}
\end{equation}
The radial velocity of the probe is obtained combining Eqs.~(\ref{eq:5.5}) and (\ref{eq:5.6})
\begin{equation}
V_r^2 = {V^2} - V_{\theta} ^2.
\label{eq:5.7}
\end{equation}
The error function becomes
\begin{equation}
\Im (a,e) = {\left({V_\theta }- {{V_S}} \right)^2} + V_r^2,
\label{eq:5.8}
\end{equation}
and its variation is
\begin{equation}
\delta V_{ex}^2 = 2\left( {{V_\theta } - {V_S}} \right)\delta {V_\theta } + 2{V_r}\delta {V_r}.
\label{eq:5.9}
\end{equation}
From Eq.~(\ref{eq:5.7}) we obtain the variation of radial velocity at intercept
\begin{equation}
{V_r}\delta {V_r} = V\delta V - {V_\theta }\delta {V_\theta },
\label{eq:5.10}
\end{equation}
where the change in velocity magnitude comes from Eq.~(\ref{eq:5.5})
\begin{equation}
2V\delta V = \frac{\mu }{{{a^2}}}\delta a.
\label{eq:5.11}
\end{equation}
Using Eq.~(\ref{eq:5.6}), we can compute the variation of circumferential velocity:
\begin{equation}
2{V_\theta }\delta {V_\theta } = \frac{\mu }{{r_S^2}}\left[ {(1 - {e^2})\delta a - 2ae\delta e} \right].
\label{eq:5.12}
\end{equation}
Combining Eqs.~(\ref{eq:5.10})-(\ref{eq:5.12}) gives
\begin{equation}
2{V_r}\delta {V_r} = \mu \left[ {\left( {\frac{1}{{{a^2}}} - \frac{{1 - {e^2}}}{{r_S^2}}} \right)\delta a + \frac{{2ae}}{{r_S^2}}\delta e} \right].
\label{eq:5.13}
\end{equation}
Finally, we obtain the gradient of the error function from Eqs.~(\ref{eq:5.9}), (\ref{eq:5.12}) and (\ref{eq:5.13}):
\begin{equation}
\delta \Im  = \displaystyle \mu \left[ {\begin{array}{*{20}{cc}}
\left({\dfrac{1}{{{a^2}}} - \dfrac{{1 - {e^2}}}{{r_S^2}}\dfrac{{{V_S}}}{{{V_\theta }}}}\right) &
{\dfrac{{2ae}}{{r_S^2}}\dfrac{{{V_S}}}{{{V_\theta }}}}
\end{array}} \right] \cdot \left[ {\begin{array}{*{20}{c}}
{\delta a}\\
{\delta e}
\end{array}} \right] = \nabla \Im^T \cdot \delta {\bf{o }}.
\label{eq:5.14}
\end{equation}
In case the trajectories of the planet and S/C do not intersect, Eq.~(\ref{eq:5.7}) gives an imaginary value for the radial velocity. This is easily fixed using
\begin{equation}
V_r^2 = \left| {{V^2} - V_\theta ^2} \right|.
\label{eq:5.15}
\end{equation}
Note that Eq.~(\ref{eq:5.15}) gives components of the velocity which are not physical, but a useful value of the error function is obtained nonetheless (it is no longer the hyperbolic excess velocity, but it decreases as the probe's trajectory tends to the planet's orbit). With this change, the error function gradient for orbits that do not intersect becomes
\begin{equation}
{\left. {\nabla \Im } \right|_{NI}} = \mu \left[ {\begin{array}{*{20}{c}}
{\dfrac{{1 - {e^2}}}{{r_S^2}}\left( {2 - \dfrac{{{V_S}}}{{{V_\theta }}}} \right) - \dfrac{1}{{{a^2}}}}\\
{\dfrac{{2ae}}{{r_S^2}}\left( {\dfrac{{{V_S}}}{{{V_\theta }}} - 2} \right)}
\end{array}} \right].
\label{eq:5.16}
\end{equation}
$V_{\theta}$ in Eqs.~(\ref{eq:5.14}) and (\ref{eq:5.16}) is always computed using Eq.~(\ref{eq:5.6}) regardless of whether the value is realistic or not.

Direct use of Eqs.~(\ref{eq:5.14}) and (\ref{eq:5.16}) leads to poor performance of the guidance algorithm. Due to the absolute value in Eq.~(\ref{eq:5.15}), the gradient of the error function presents a discontinuity when the orbits of the S/C and planet are tangent (i.e., when the aphelion $r_{\alpha}$ of the probe equals the orbital radius of the planet). The error surface, when visualized in 3D, presents a narrow valley with a sharp edge at the bottom. The sudden change in slope forces the steepest descent algorithm into a zigzag path along the line of discontinuity, degrading the performance. This is a common problem with steepest descent methods. It can be mitigated with an inertia term that smooths changes in the direction of motion. This gives rise to the class of nonlinear conjugate gradient methods \cite{fletcher1964,polak1969,hestenes1952}. In our specific application, the inertia would affect the thrust angle. However, there is a much more efficient alternative because the correct path to follow when the gradient discontinuity problem arises can be determined beforehand. It is the curve
\begin{equation}
r_{\alpha} = a(1 + e) = {r_S},
\label{eq:5.17}
\end{equation}
that is, the orbital elements of the S/C must evolve in a way that keeps the aphelion constant. To this end, once the probe's osculating Keplerian trajectory becomes tangent to Saturn's orbit, the guidance algorithm switches to aphelion hold mode. The thrust direction is chosen such that, at the end of the current time step, the predicted aphelion coincides with the orbit of the planet:
\begin{equation}
{r_{\alpha}} + \frac{{{\rm{d}}{r_{\alpha}}}}{{{\rm{d}}t}}\Delta t = {r_S},
\label{eq:5.18}
\end{equation}
where $\Delta t$ is the time step of the numerical integrator. We can rearrange Eq.~(\ref{eq:5.18}) as
\begin{equation}
\frac{{{\rm{d}}{r_{\alpha}}}}{{{\rm{d}}t}} = \frac{{{r_S} - {r_{\alpha}}}}{{\Delta t}} = {\bar{\dot{r}}_{\alpha}},
\label{eq:5.19}
\end{equation}
which expands into
\begin{equation}
\nabla {r_{\alpha}}^T \cdot \frac{{{\rm{d}}{\bf{o}}}}{{{\rm{d}}t}} = {\bar{\dot{r}}_{\alpha}}\,\,{\rm{   where   }}\,\,\nabla {r_{\alpha}} = \left[ {\begin{array}{*{20}{c}}
{1 + e}\\
a
\end{array}} \right].
\label{eq:5.20}
\end{equation}
Using Eq.~(\ref{eq:3.5})
\begin{equation}
\nabla {r_{\alpha}}^T \cdot {\bf{R}} \cdot {\bf{f}} = {\bar{\dot{r}}_{\alpha}}.
\label{eq:5.21}
\end{equation}
Let $\nabla {r_{\alpha}}^T \cdot {\bf{R}} = \left[ {\begin{array}{*{20}{c}}
{{d_1}}&{{d_2}}
\end{array}} \right]$. Upon substituting Eq.~(\ref{eq:3.1}), Eq.~(\ref{eq:5.21}) becomes
\begin{equation}
{d_1}\sin \beta  + {d_2}\cos \beta  = \frac{\bar{\dot{r}}_{\alpha}}{f_{\max }} = {d_3}.
\label{eq:5.22}
\end{equation}
To solve this equation, let us define
\begin{equation}
l = \sqrt {d_1^2 + d_1^2} \,\,\,;\,\,\,\varphi = \arctan \frac{{{d_1}}}{{{d_2}}},
\label{eq:5.23}
\end{equation}
such that
\begin{equation}
{d_1} = l\sin \varphi \,\,\,;\,\,\,{d_2} = l\cos \varphi.
\label{eq:5.24}
\end{equation}
Substituting in Eq.~(\ref{eq:5.22})
\begin{equation}
l\left( {\sin \varphi \sin \beta  + \cos \varphi \cos \beta } \right) = {d_3}.
\label{eq:5.25}
\end{equation}
Thus
\begin{equation}
\cos \left( {\varphi  - \beta } \right) = \frac{{{d_3}}}{l}\,\,\, \to \,\,\,\beta  = \varphi  \pm \arccos \left( {\frac{{{d_3}}}{l}} \right).
\label{eq:5.26}
\end{equation}
From the two solutions\footnote{Clearly, Eq.~(\ref{eq:5.26}) is not valid if $\left| {{d_3}} \right| > l$. This means the aphelion error is too large to be corrected in a single time step. In that case, use $\beta = \varphi$ and proceed normally.} of Eq.~(\ref{eq:5.26}), choose the value of $\beta$ that causes the fastest rate of decrease of the excess velocity. Let $\beta_1$ be the correct choice, then
\begin{equation}
{\left. {\frac{{{\rm{d}}V_{ex}^2}}{{{\rm{d}}t}}} \right|_{{\beta _1}}} < {\left. {\frac{{{\rm{d}}V_{ex}^2}}{{{\rm{d}}t}}} \right|_{{\beta _2}}}.
\label{eq_5.27}
\end{equation}

\section{Choice of threshold excess velocity}
\label{sec:choice}
\begin{figure}
	\centering\includegraphics[width=8cm]{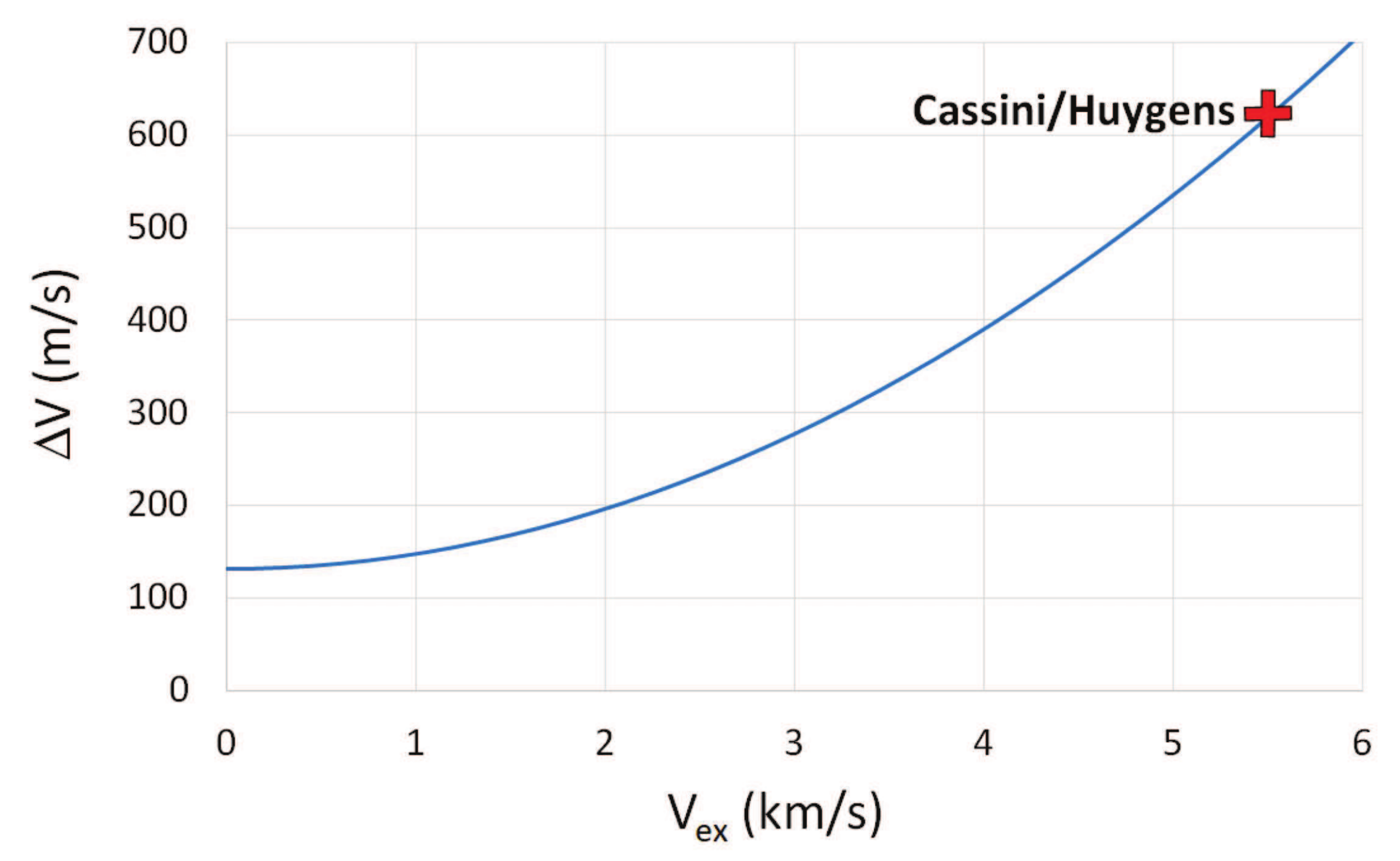}
	\caption{Saturn orbit insertion impulse vs. hyperbolic excess velocity. Cassini/Huygens mission shown for reference.}
	\label{fig:dv_vex}
\end{figure}

Lowering the hyperbolic excess velocity at Saturn reduces the amount of propellant required for the orbit insertion burn. However, there is a point where further reductions of $V_{ex}$ have a negligible impact on the capture maneuver. Therefore, a reasonable lower bound $V_{th}$ must be established at which the electric propulsion system is deactivated and the S/C continues in a ballistic trajectory. We based our analysis on the initial orbit (i.e., immediately after the OI burn) of the Cassini/Huygens mission. This trajectory had a pericenter radius of 80230 km and an orbital period of 120 days. Assuming an impulsive burn at the pericenter, we computed the $\Delta V$ required for OI as a function of the hyperbolic excess velocity. Figure~\ref{fig:dv_vex} plots the OI impulse variation with $V_{ex}$; the marker shows the Cassini mission for reference purposes. Values of $V_{ex}$ below 1 km/s have  little impact on $\Delta V$. This happens because the contribution of the excess velocity to the total planetocentric energy of the S/C becomes small compared to the changes in potential energy. Thus, the magnitude of the impulse is almost the same as in the case of a parabolic orbit (132 m/s). For this reason, we have chosen a threshold velocity $V_{th}$ = 1 km/s yielding a $\Delta V$ of 148 m/s (just 12\% above the absolute minimum). The orbit propagator monitors the value of $V_{ex}$ to determine if it falls below $V_{th}$. When that happens, the orbital elements are frozen and Keplerian motion is assumed until the encounter with Saturn.

Note that, once the excess velocity has been reduced, the OI burn can be avoided altogether if a sufficiently deep (i.e., 1000 km altitude, the upper limit of the atmosphere) gravity assist with Titan is performed upon arrival. The period of the resulting elliptical orbit can be as low as 78 days, with a pericenter altitude of $\sim$1 million km. In contrast, a purely ballistic (i.e., unpowered) Jupiter-to-Saturn trajectory  yields a minimum $V_{ex}$ of 2.5 km/s ($\sim$1 km/s higher than in a Jupiter-Saturn Hohmann transfer). It is still possible to achieve capture with a Titan flyby, but the period of the final orbit would be over 8 months, with very large eccentricity. Such a trajectory would be far less convenient for scientific activities. The methods and technologies to execute the OI will be addressed in a future continuation of the work.

\section{Results and discussion}
\label{sec:discuss}

\subsection{Choice of the parameters}
Before discussing the best trajectories obtained, we shall briefly analyze the range of free parameters (launch $C_3$, initial flight path angle and flyby depth during GA) that yield acceptable solutions. We shall also study their effect on the mission characteristics (e.g., transfer duration and OI impulse required). As a recap, the constraints imposed on the trajectories where:
\begin{itemize}
	\item Earth-to-Jupiter transfer time under 3 years
	\item Maximum duration of post-GA thrust arc of 4 years
	\item Value of hyperbolic excess velocity at thruster cutoff of 1 km/s
\end{itemize} 
Additionally, to avoid overwhelming the reader with data of little relevance, we have filtered out solutions yielding large values of the hyperbolic excess velocity on arrival at Saturn. We have set the upper limit at 1.3 km/s, which corresponds to an orbit insertion $\Delta V$ of 159 m/s (20\% more than the parabolic limit).

In order to maximize the S/C mass and/or reduce the requirements of the launcher, $C_3$ must be kept near its lower limit. Under these conditions, as we discussed previously, the compatible $\gamma$ values are small and the departure trajectories are almost tangent to Earth's orbit. Therefore, the most relevant parameters are $C_3$ and the fly-by depth, as the initial flight path angle is constrained to a very narrow range. 

\begin{figure}
	\centering\includegraphics[width=8cm]{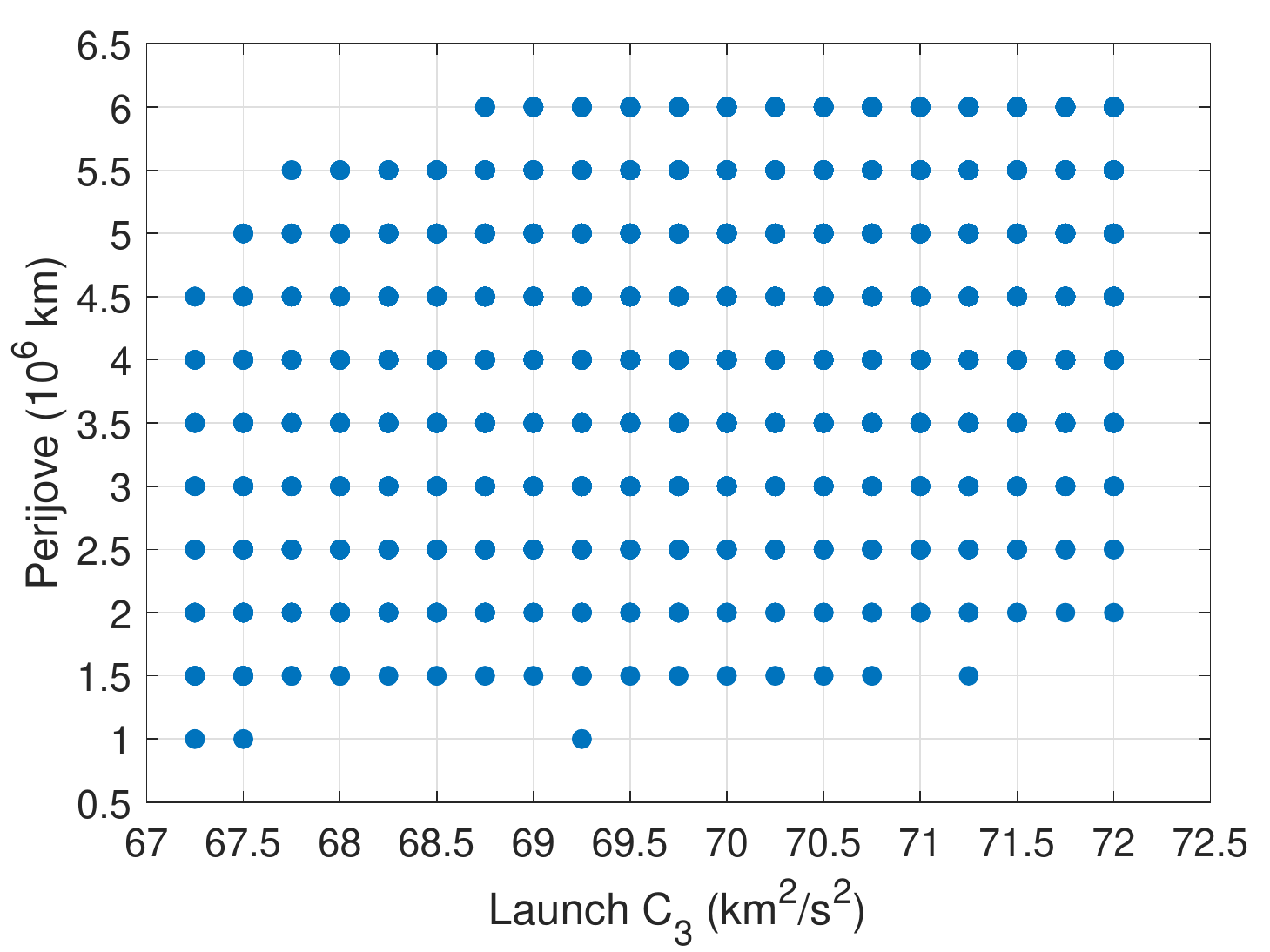}
	\caption{Combinations of launch $C_3$ and fly-by depth that respect the mission constraints.}
	\label{fig:rjupi_c30}
\end{figure}

Figure~\ref{fig:rjupi_c30} shows the combinations of $C_3$ and fly-by depth that fulfill our requirements. Keep in mind that each marker on the plot may correspond to more than one value of $\gamma$, specially for the higher values of $C_3$.
Trajectories with a high perijove (above 4.5 to 6 million kilometers, depending on launch energy) give rise to post-GA ellipses with moderate eccentricity and semimajor axis. These do not intersect Saturn's orbit and a long thrust arc is needed just to reach the destination planet. As the thrust duration is limited, the resulting trajectories fail to reach Saturn, or do so with a large relative velocity.
On the other end of the spectrum, when the minimum distance to Jupiter is small (below 1 to 2 million kilometers, depending on $C_3$) the post-GA ellipses show high values of the aphelion distance and eccentricity. They intersect Saturn's orbit, but the initial relative velocity is so high that it cannot be reduced to acceptable levels before the thruster is powered down. In fact, in the most extreme cases, the S/C exceeds the escape velocity of the solar system, resulting in a hyperbolic trajectory.

The results show that it is possible to achieve the threshold hyperbolic excess velocity (1 km/s) even for the smallest departure energy $C_3 =$ 67.25 (km/s)$^{2}$. Therefore, the real constraint for $C_3$ is not the excess velocity, but being able to reach Jupiter in less than 3 years.

\subsection{Jupiter-to-Saturn transfer phases}
\begin{figure}
	\centering\includegraphics[width=9cm]{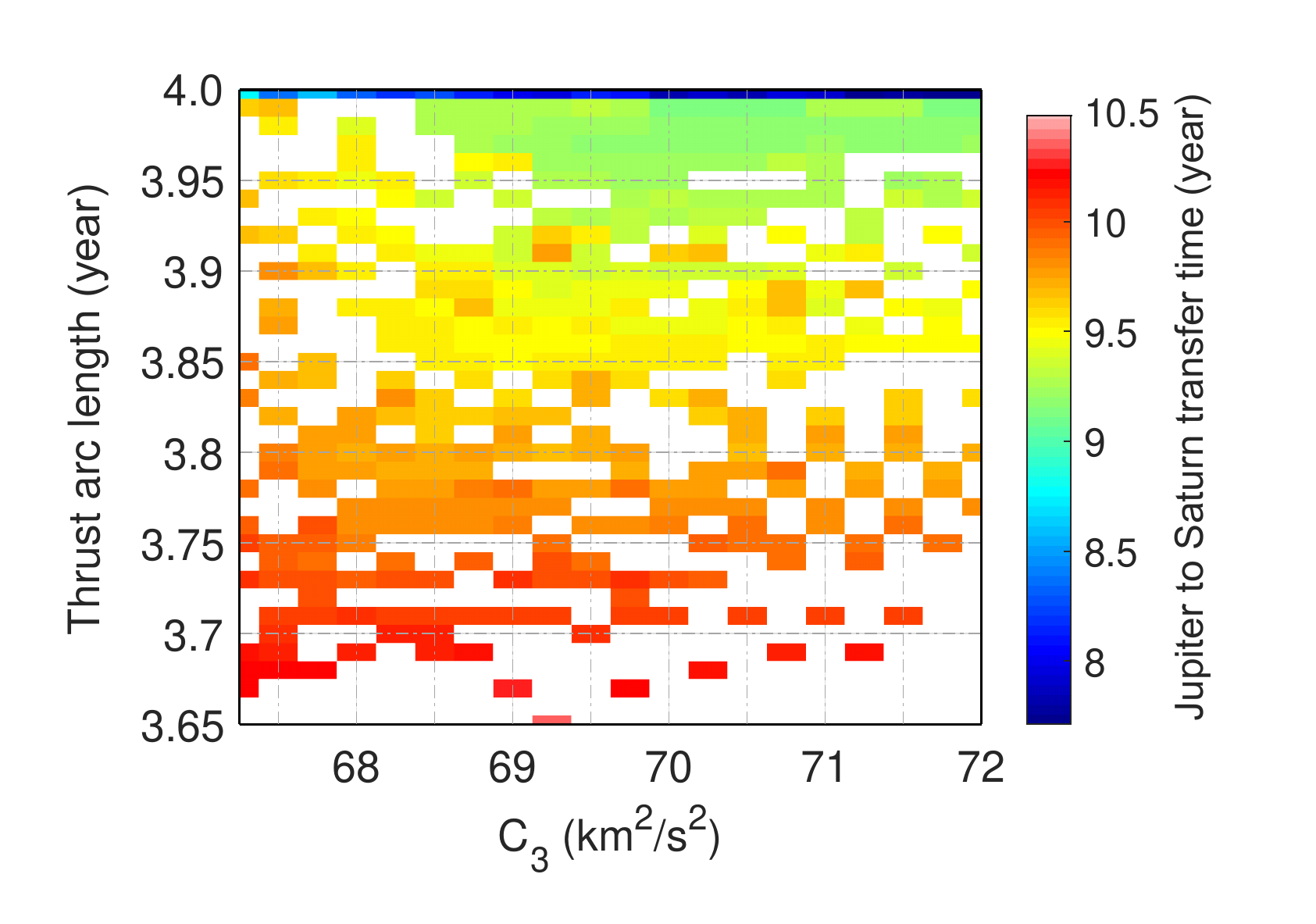}
	\caption{Jupiter-Saturn transfer time  vs. post-GA thrust arc length and launch $C_3$.}
	\label{fig:t23_c30}
\end{figure}

The consumption of propellant during the Jupiter-to-Saturn segment is directly proportional to the duration of the thrust arc (we assume a constant propellant flow rate, in spite of keeping the acceleration constant, which is a conservative approach). Therefore, trajectories with the shortest thrust arc are desirable. Figure~\ref{fig:t23_c30} shows the combinations of thrust duration and $C_3$ that meet the mission requirements and the associated total Jupiter-to-Saturn transfer time, including powered and coast phases. The upper part of the chart (4 years of thrust) correspond to excess velocities higher than 1 km/s (and below 1.3 km/s). The rest of the trajectories all have relative velocities of 1 km/s (because the thrust has been deactivated before the 4 years mark). The chart shows that minimum thrust arc duration depends weakly on $C_3$, and is always very close to 3.67 years, even for the minimum departure energy. This translates into 209 kg of propellant for the Jupiter-to-Saturn transfer (20\% of the initial S/C mass). Note that the difference between the best and worst solutions in terms of thrust duration is only 0.33 years, equivalent to less than 19 kg of propellant (under 2\% of the initial S/C mass). Therefore, in terms of mass budget, all the trajectories are comparable.

\begin{figure}
	\centering\includegraphics[width=8cm]{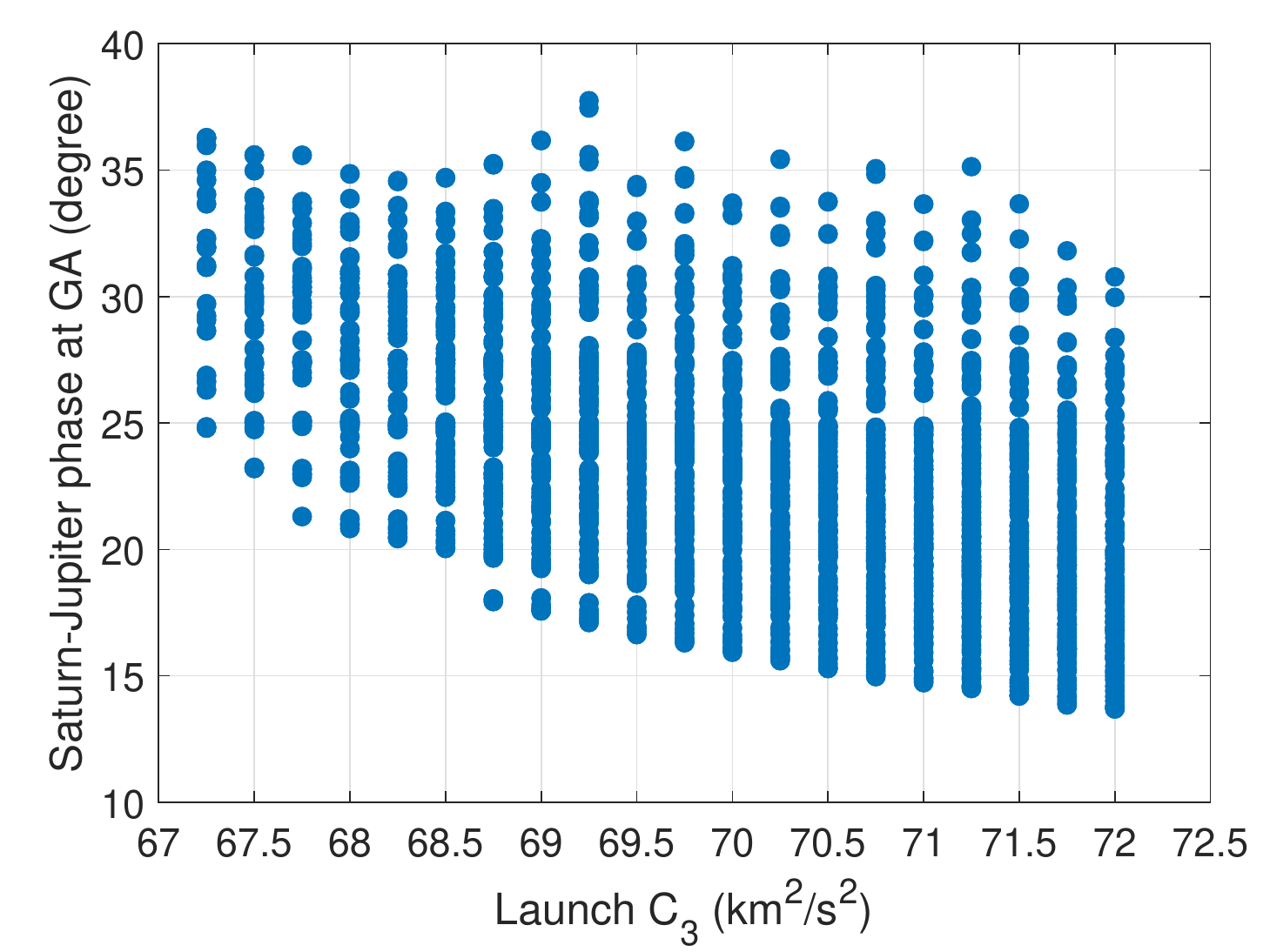}
	\caption{Saturn-Jupiter relative phase at the time of GA vs. launch $C_3$.}
	\label{fig:phase_c30}
\end{figure}

The trajectories with minimal thrust duration have a Jupiter-to-Saturn transfer time close to 10 years irrespective of $C_3$. As the thrust arc approaches 4 years, the transfer time decreases to 9 years. The trajectories in the upper edge of Fig.~\ref{fig:t23_c30} (4 years of thrust) can shorten the transfer to 7.7 years. However, this comes at the expense of higher hyperbolic excess velocity (due to higher eccentricity). The differences in transfer duration arise from the coast phase, because the variations in thrust arc length are comparatively small (0.3 years).
The long duration of this transfer is undesirable, but it is an inescapable consequence of the mission goal. In order to minimize $V_{ex}$, the trajectory of the probe must be tangent to Saturn's orbit and as close to circular as possible. This makes to Jupiter-to-Saturn segment very long by design. The duration of this phase determines the launch window opportunities because the synodic period of the Earth-Jupiter system (very close to 1 year) is much smaller than for the Jupiter-Saturn pair (almost 20 years). Figure~\ref{fig:phase_c30} presents the relative planetary phase at the time of Jupiter GA (i.e., how many degrees Saturn leads Jupiter in the heliocentric reference frame at the time of fly-by). The phase angle ranges between 14 and 38 degrees, which means the window of opportunity remains open for 16 months. At the time of writing, the window is open and will close in mid-February 2020. The next opportunity will start in August 2038. $C_3$ has a moderate effect on window duration, it becomes 10\% shorter at low launch energies.

\subsection{Earth-to-Saturn transfer time and propellant budget}
\begin{figure}
	\centering\includegraphics[width=8cm]{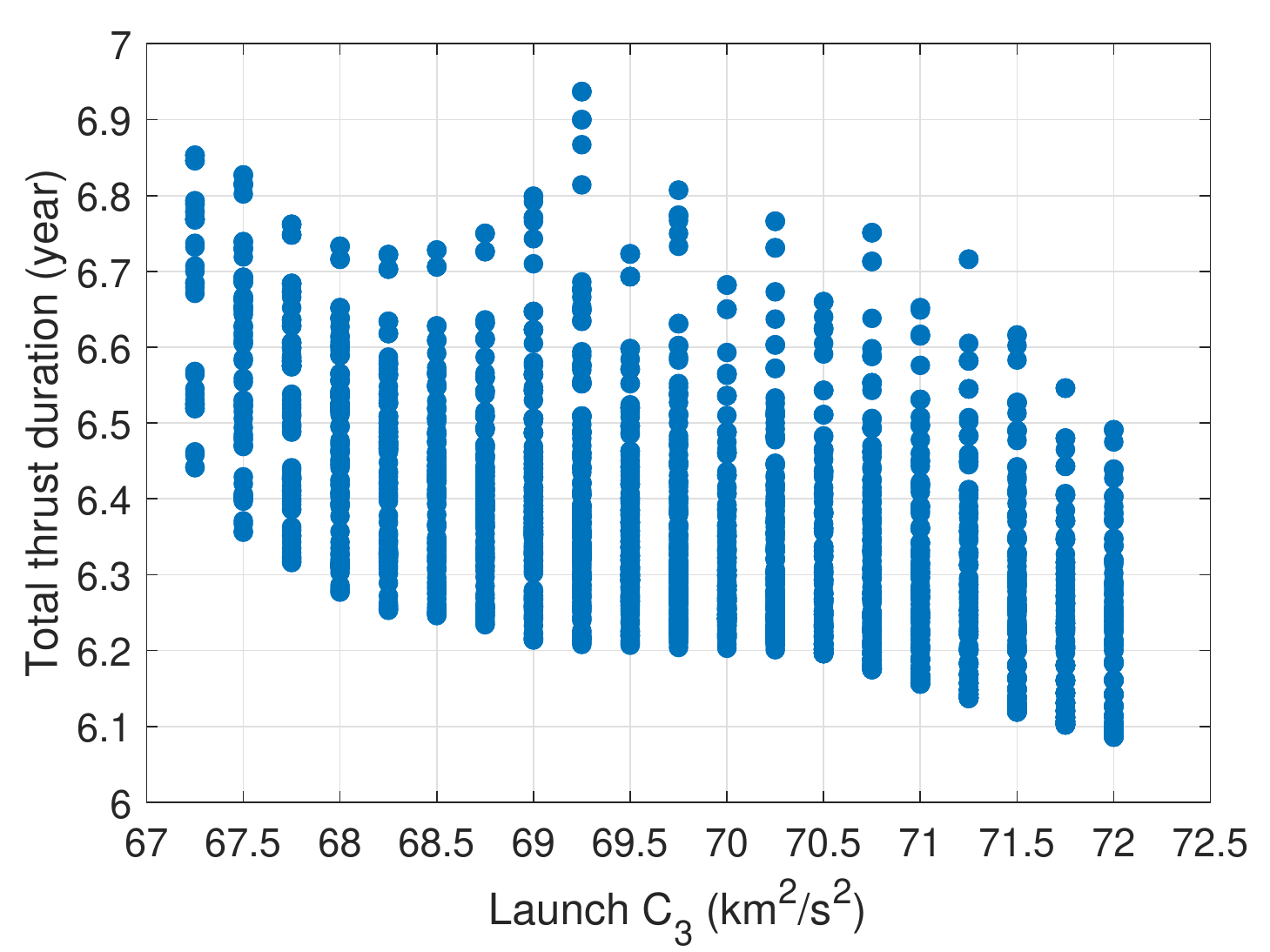}
	\caption{Total length of the powered phase vs. launch $C_3$.}
	\label{fig:t03_c30}
\end{figure}

We conclude the preliminary analysis presenting the total duration of the powered phase (i.e., including the pre- and post-GA thrust arcs) which determines the total propellant budget. Figure~\ref{fig:t03_c30} shows that the minimum thrust duration ranges from 6.4 years at the lowest $C_3$ to 6.1 for high departure energy. The variation is mostly due to the Earth-to-Jupiter segment. Again, for the sake of economy, it is preferable to reduce the launch energy at the expense of a slight increase in electric propellant consumption. 

We have established that the most efficient trajectories are obtained using the minimum launch energy. Moving forward, we shall restrict the discussions to solutions with $C_3$=67.25 (km/s)$^{2}$. In addition to being representative of the most economically attractive case, eliminating one free parameter will reduce clutter in the presentation of the results.

\begin{figure}
	\centering\includegraphics[width=8cm]{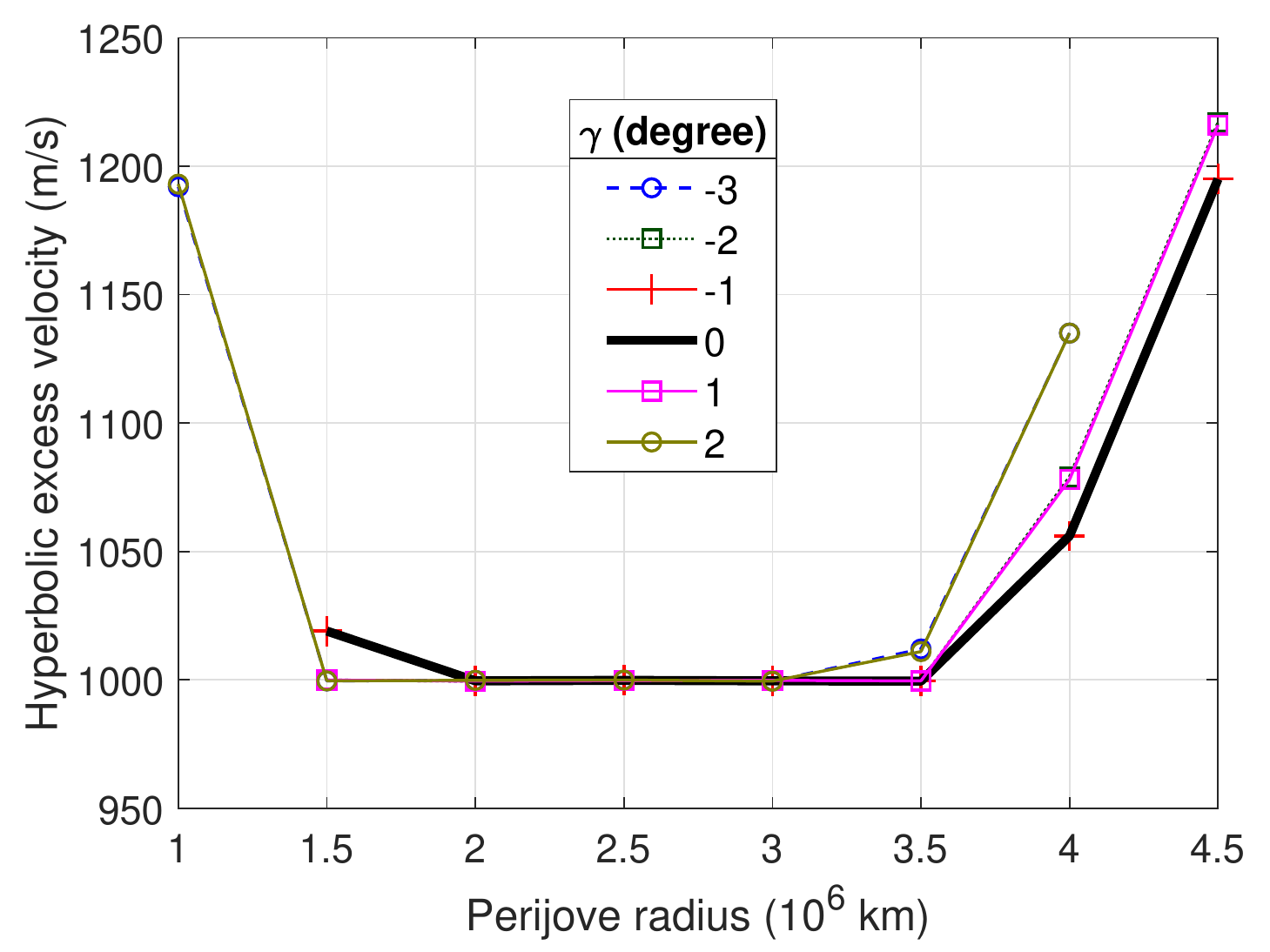}
	\caption{Saturn $V_{ex}$ vs. fly-by depth for launch $C_3$=67.25 (km/s)$^{2}$.}
	\label{fig:low_vex_rpi}
\end{figure}

Figure~\ref{fig:low_vex_rpi} plots the hyperbolic excess velocity  as a function of the fly-by depth for different values of the initial flight path angle (and $C_3$=67.25 (km/s)$^{2}$). It is possible to achieve the minimum relative velocity for depths between 1.5 and 3.5 million kilometers.

\begin{figure}
	\centering\includegraphics[width=8cm]{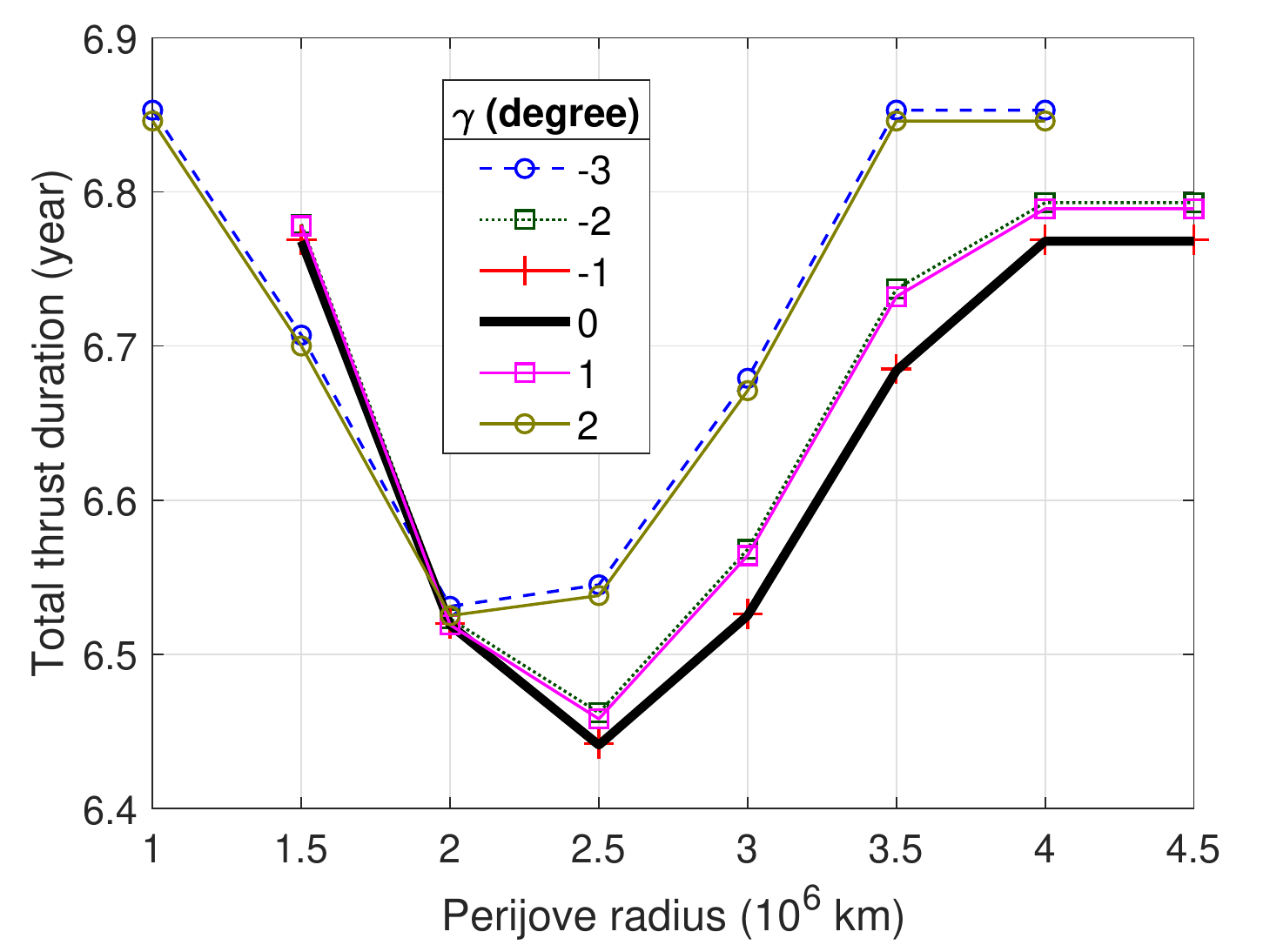}
	\caption{Total length of the powered phase vs. fly-by depth for launch $C_3$=67.25 (km/s)$^{2}$.}
	\label{fig:low_t03_rpi}
\end{figure}

\begin{figure}
	\centering\includegraphics[width=8cm]{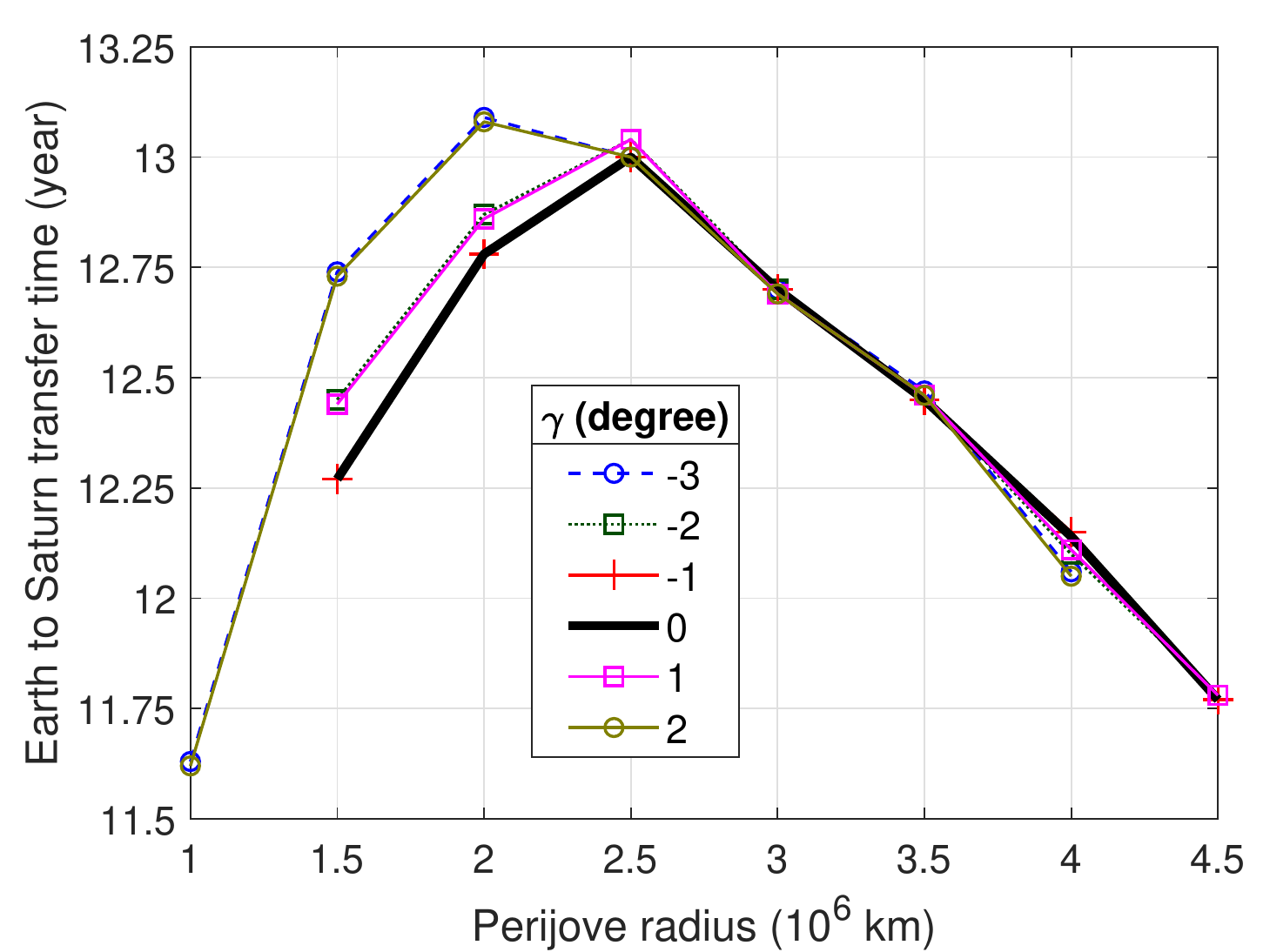}
	\caption{Earth-to-Saturn transfer time vs. fly-by depth for launch $C_3$=67.25 (km/s)$^{2}$.}
	\label{fig:low_t04_rpi}
\end{figure}

The corresponding duration of the propelled phase (including pre- and post-GA thrust arcs) is depicted in Figure~\ref{fig:low_t03_rpi}. All the trajectories that yield the minimum excess velocity have comparable total thrust duration, ranging from 6.44 to 6.85 years. Their Earth-to-Saturn transfer time varies between 12.3 and 13.1 years (see Figure~\ref{fig:low_t04_rpi}). Note that the results for $\gamma$=-3$^{\circ}$ and $\gamma$=2$^{\circ}$ are virtually identical. The same goes for the flight path angle pairs (-2$^{\circ}$,1$^{\circ}$) and (-1$^{\circ}$,0$^{\circ}$).

\subsection{Solutions}
\begin{figure}
	\centering\includegraphics[width=8cm]{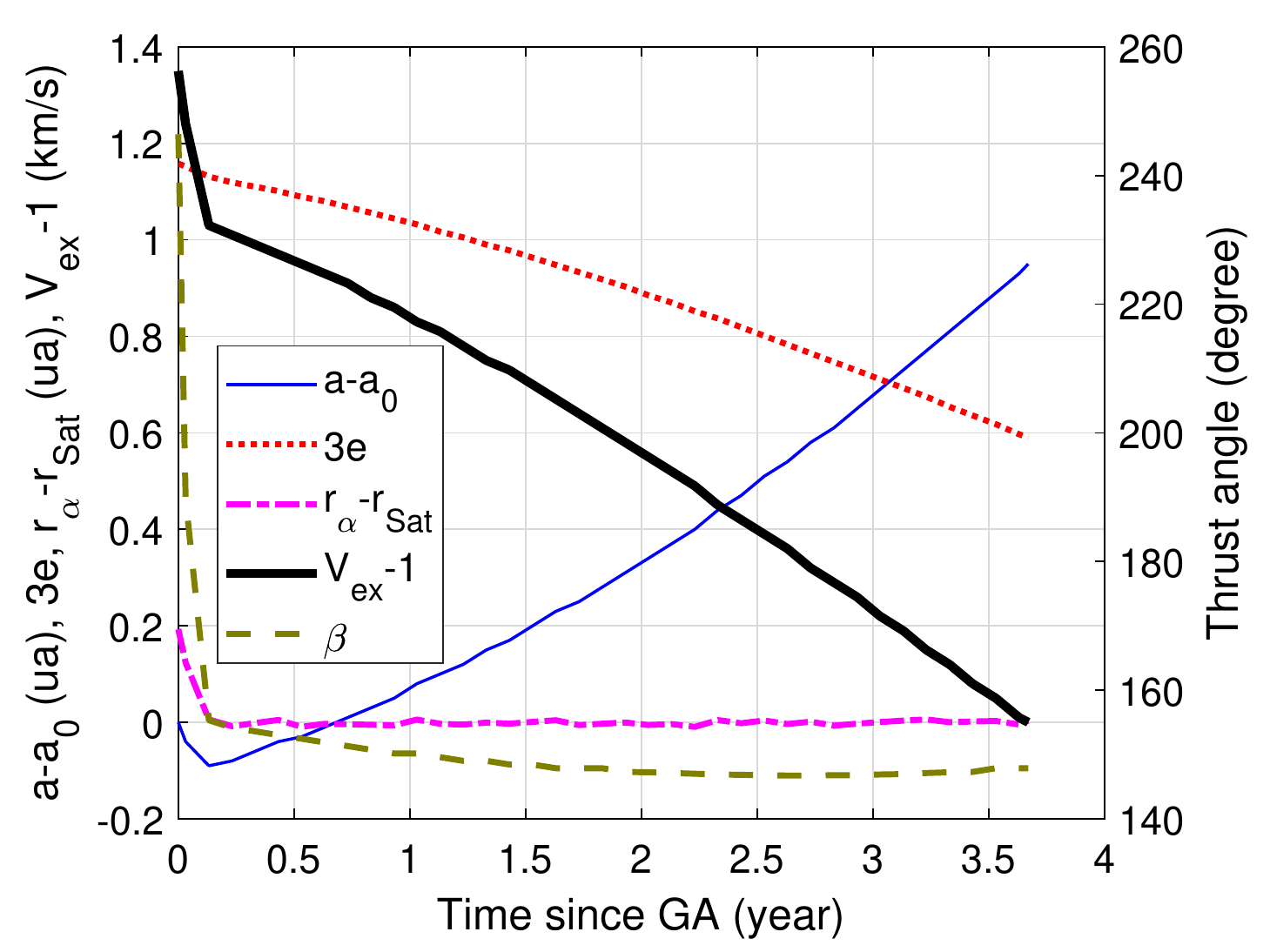}
	\caption{Orbital parameters and thrust angle vs. time after GA for the best solution ($\gamma$=0, $r_\pi$=2.5 $\cdot 10^{6}$ km, $C_3$=67.25 (km/s)$^{2}$).}
	\label{fig:case_879}
\end{figure}

\begin{figure}
	\centering\includegraphics[width=8cm]{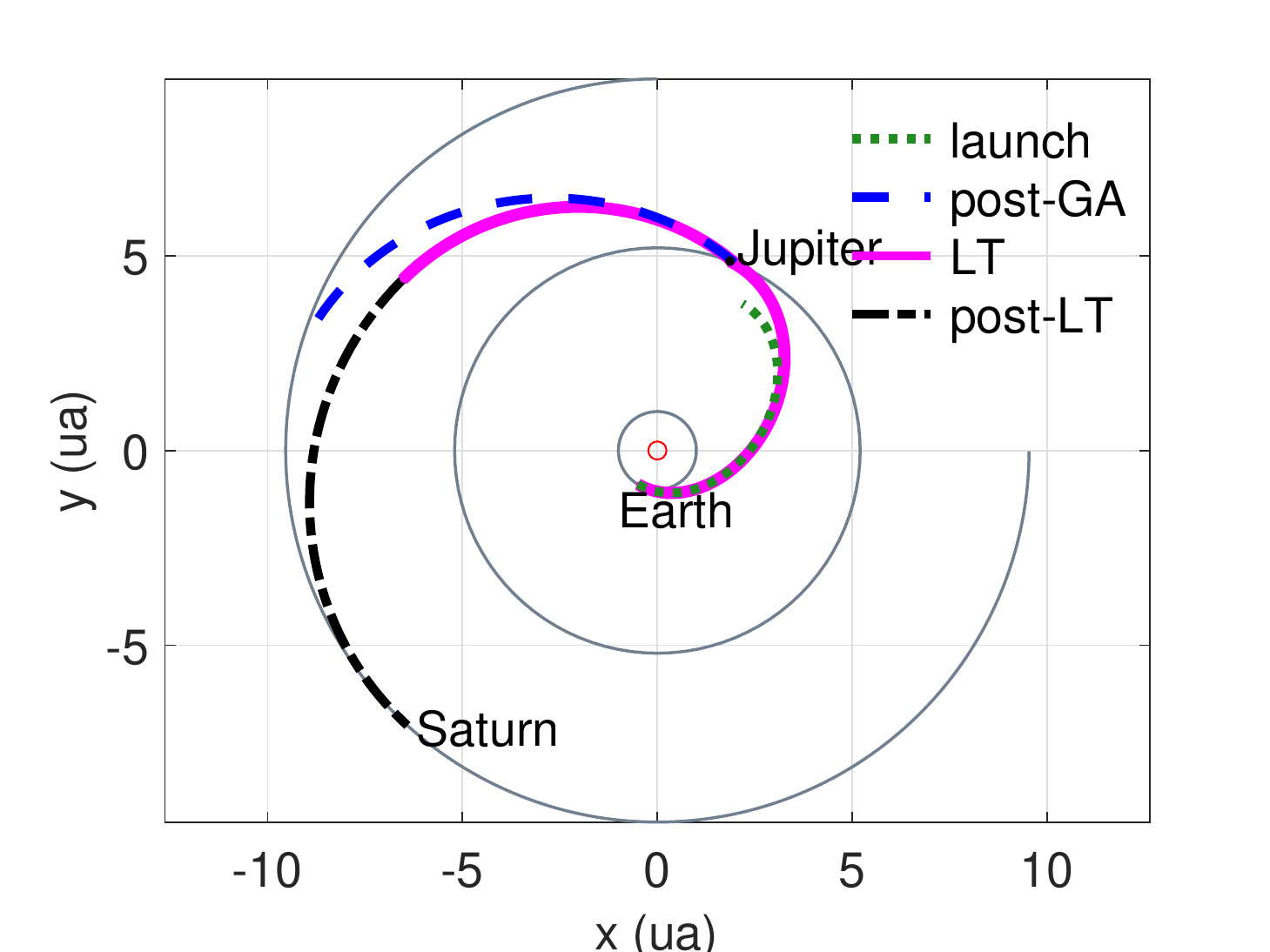}
	\caption{Powered phase (solid pink), coast segment (dot-dashed black), post-launch osculating ellipse (dotted green) and post-GA osculating ellipse (dashed blue) for the best solution.}
	\label{fig:traj_879}
\end{figure}

The best solution (minimum propellant budget) corresponds to $\gamma$=0 and $r_\pi =2.5 \cdot 10^{6}$ km. The total propellant consumption is 367 kg, split into 158 kg for reaching Jupiter and 209 kg for $V_{ex}$ reduction. The total duration of the transfer is 13 years. The orbital elements immediately after the GA are $a_0$=7.02 au and $e_0$=0.386. Thus, the initial aphelion is 9.73 au which is close to the orbital radius of Saturn $r_{S}$=9.54 au (i.e., the post-GA heliocentric ellipse is almost tangent to Saturn's orbit). Figure~\ref{fig:case_879} shows the evolution of the semi-major axis, eccentricity, aphelion, excess velocity and thrust angle against time; the complete Earth-Jupiter-Saturn trajectory is plotted in Figure~\ref{fig:traj_879}. To accommodate the variables to the scale of Figure~\ref{fig:case_879}, some have been offset or scaled:
\begin{itemize}
	\item The semi-major axis has been offset by its value immediately after the encounter with Jupiter ($a_0$)
	\item The eccentricity $e$ has been scaled by a factor of 3
	\item The aphelion $r_\alpha$ has been offset by the orbital radius $r_{S}$ of Saturn. The guidance algorithm enters aphelion hold mode when this parameter becomes null
	\item The hyperbolic excess velocity $V_{ex}$ has been offset by its target value (1 km/s). The engine is shut down when this parameter reaches zero
\end{itemize} 

There are two distinct phases of the thrust arc. Initially, when the aphelion is above the orbit of Saturn, the fastest way of lowering $V_{ex}$ is reducing the semi-major axis until the two orbits become tangent. The thrust angle starts at 246$^{\circ}$, which corresponds to a strong braking action. This is a very effective way of lowering the aphelion, so this phase  spans only a small part (50 days) of the thrust arc. In reality, the eccentricity is also slightly reduced during this period, but its effect on the aphelion is minor.

The second stage lasts much longer (3.54 years) and is spent circularizing the trajectory while maintaining the aphelion fixed. This phase is characterized by a steady decrease of the eccentricity, which goes from 0.386 after the GA to 0.197 at the time of thruster cutoff. Meanwhile, the thrust angle $\beta$ remains close to 150$^{\circ}$ providing positive mechanical power which increases the semi-major axis at a rate that balances the eccentricity reduction. During the second stage, the rate of reduction of the excess velocity is almost an order of magnitude smaller (0.29 km/s per year on average) than during the initial phase (2.42 km/s per year). This means the sensitivity of $V_{ex}$ to eccentricity is much smaller than to $r_\pi$. The final orbital elements (i.e., when the coast phase starts) are $a_f$=7.97 au and $e_f$=0.197.

\begin{figure}
	\centering\includegraphics[width=8cm]{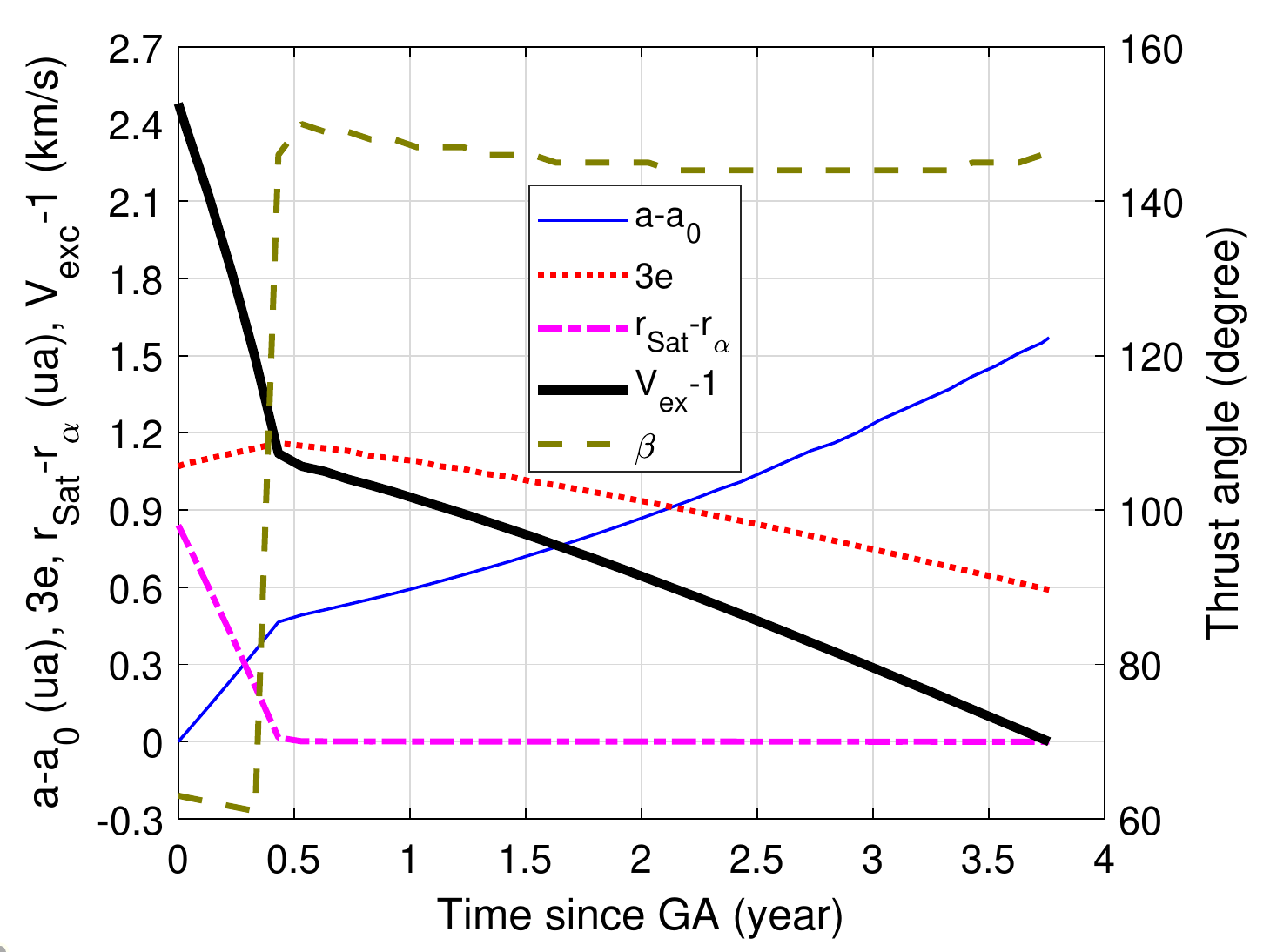}
	\caption{Orbital parameters and thrust angle vs. time after GA for trajectory with $\gamma$=0, $r_\pi$=3 $\cdot 10^{6}$ km, $C_3$=67.25 $(km/s)^{2}$.}
	\label{fig:case_880}
\end{figure}

\begin{figure}
	\centering\includegraphics[width=8cm]{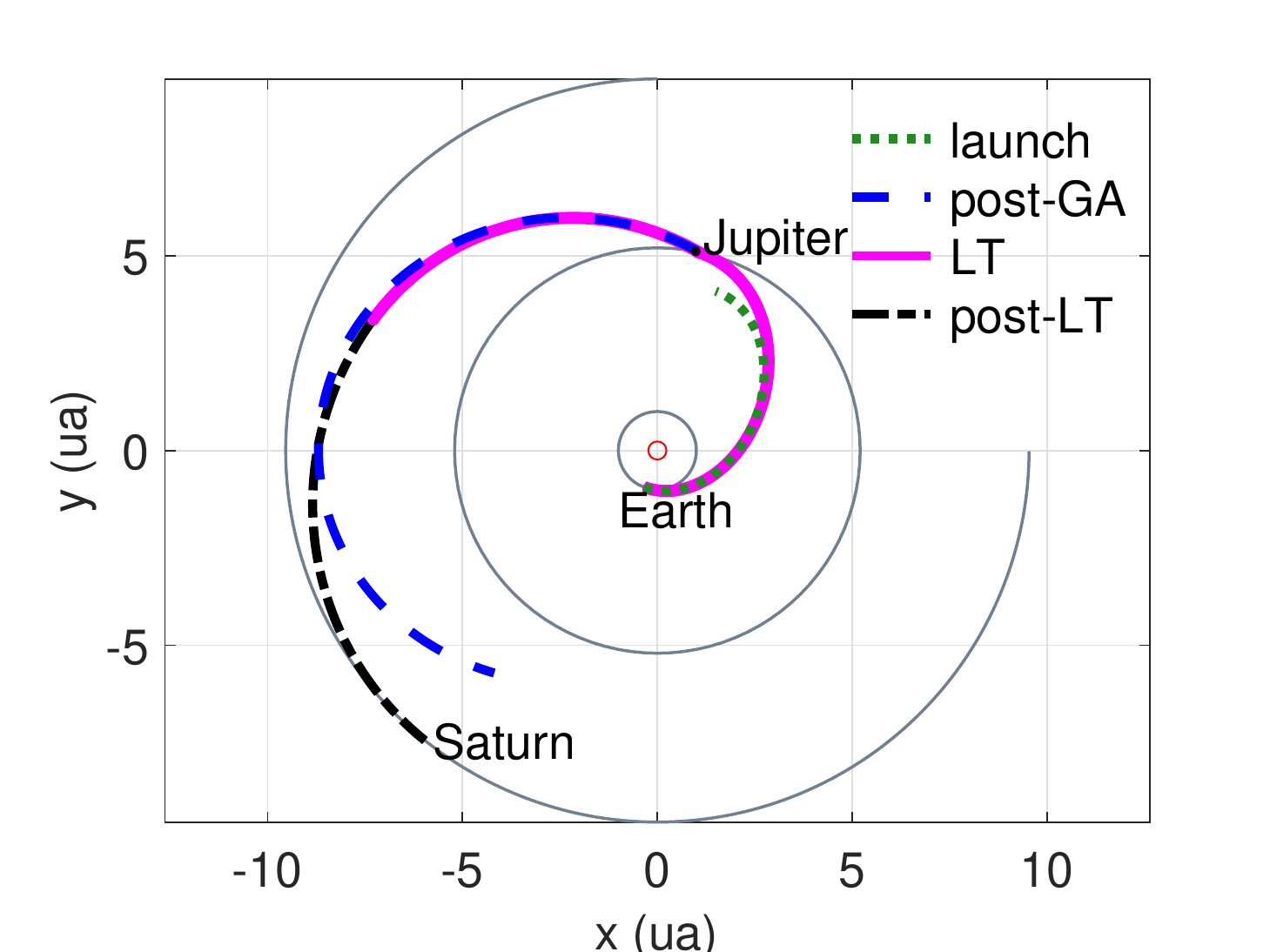}
	\caption{Powered phase (solid pink), coast segment (dot-dashed black), post-launch osculating ellipse (dotted green) and post-GA osculating ellipse (dashed blue) for trajectory with $\gamma$=0, $r_\pi$=3 $\cdot 10^{6}$ km, $C_3$=67.25 (km/s)$^{2}$.}
	\label{fig:traj_880}
\end{figure}

To illustrate a different type of trajectory, Figures~\ref{fig:case_880} and~\ref{fig:traj_880} display a solution very similar to the best case, but with a larger fly-by radius ($r_\pi =3 \cdot 10^{6}$ km). The post-GA orbital elements are now $a_0$=6.40 au and $e_0$=0.358. Due to the weaker GA boost, the initial aphelion is 8.69 au; short of Saturn's orbit by 0.85 au. Note that this time we plot $r_{S}-r_{\alpha}$ to keep the value positive.
Again, we observe two stages. Initially, when the osculating ellipse does not intersect Saturn's orbit, the control law raises the aphelion as quickly as possible. $\beta$ averages 62$^{\circ}$, which corresponds to a thrust vector almost parallel to the S/C velocity. This yields the maximum mechanical power and, consequently, the fastest rate of increase of the semi-major axis. This validates our decision to use tangential thrust during the Earth-to-Jupiter leg to reach the giant planet's orbit in a short time. The reader should keep in mind that the value of $V_{ex}$ computed when $r_\pi < r_{S}$ is not physically meaningful, it serves only as error signal for the guidance algorithm. The initial stage lasts for 160 days in this case. The longer duration is due to the larger difference between the initial aphelion and Saturn's orbit. Once the control law switches to aphelion hold, the evolution is very similar to the previous example. The duration of this segment is 3.30 years and the final state is $a_f$=7.97 au and $e_f$=0.197. Note that these are exactly the same orbital elements as in the {\ best} solution. This is not surprising as, by design, both trajectories are tangent to Saturn's orbit and have the same hyperbolic excess velocity (1 km/s). Thus, the eccentricities and semi-major axes must coincide. The different post-GA conditions influence the evolution during the powered segment, but do not affect the final result (as long as $V_{ex}$ can reach the target value before the motor is shut down). However, the initial conditions do affect the duration of the thrust arc, which is longer in the second case (3.76 vs. 3.67 years) due to the lower energy imparted by the GA.  

The time of flight from Earth to Saturn is 13.0 years for the best trajectory and 12.7 for the second example. Compared to the total duration of the mission, the difference is negligible. Furthermore, the variation in propellant consumption is just 5 kg (less than $1.4\%$ of the total budget), indicating the propellant mass has a smooth minimum in the space of mission parameters. Figure~\ref{fig:low_t03_rpi} illustrates this fact, the differences in thrust duration for all solutions giving the minimum $V_{ex}$ are below 0.4 years (equivalent to 23 kg of propellant). This provides the mission planner flexibility to match additional requirements. 

\section{Comparison with a trajectory optimizer}
\label{sec:glob_loc}
\begin{figure}
	\centering\includegraphics[width=8cm]{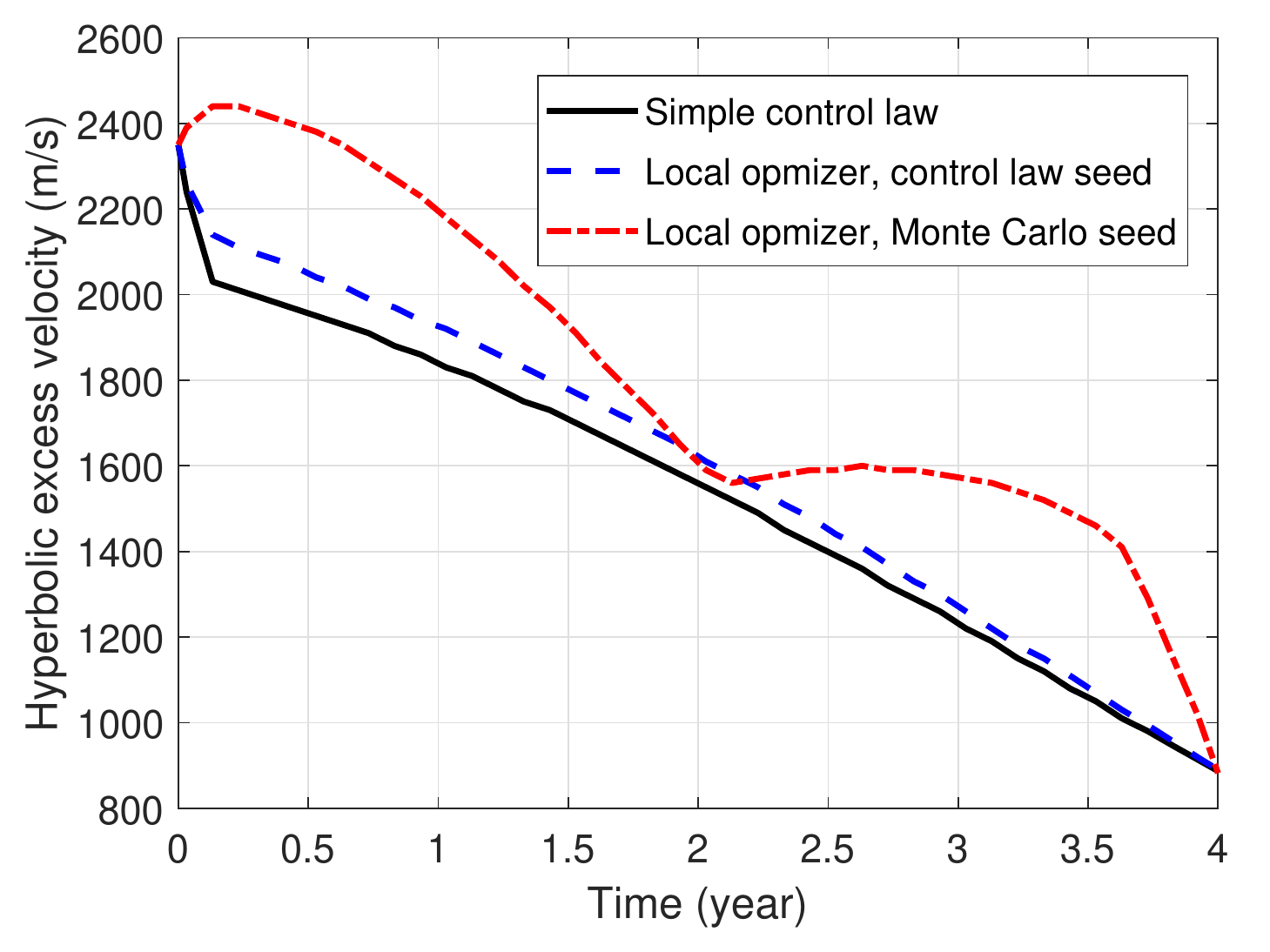}
	\caption{Excess velocity vs. time after GA obtained with steering law (solid line) and optimizer (dashed lines) ($\gamma$=0, $r_\pi$=2.5 $\cdot 10^{6}$ km, $C_3$=67.25 (km/s)$^{2}$).}
	\label{fig:global_vinf}
\end{figure}

\begin{figure}
	\centering\includegraphics[width=8cm]{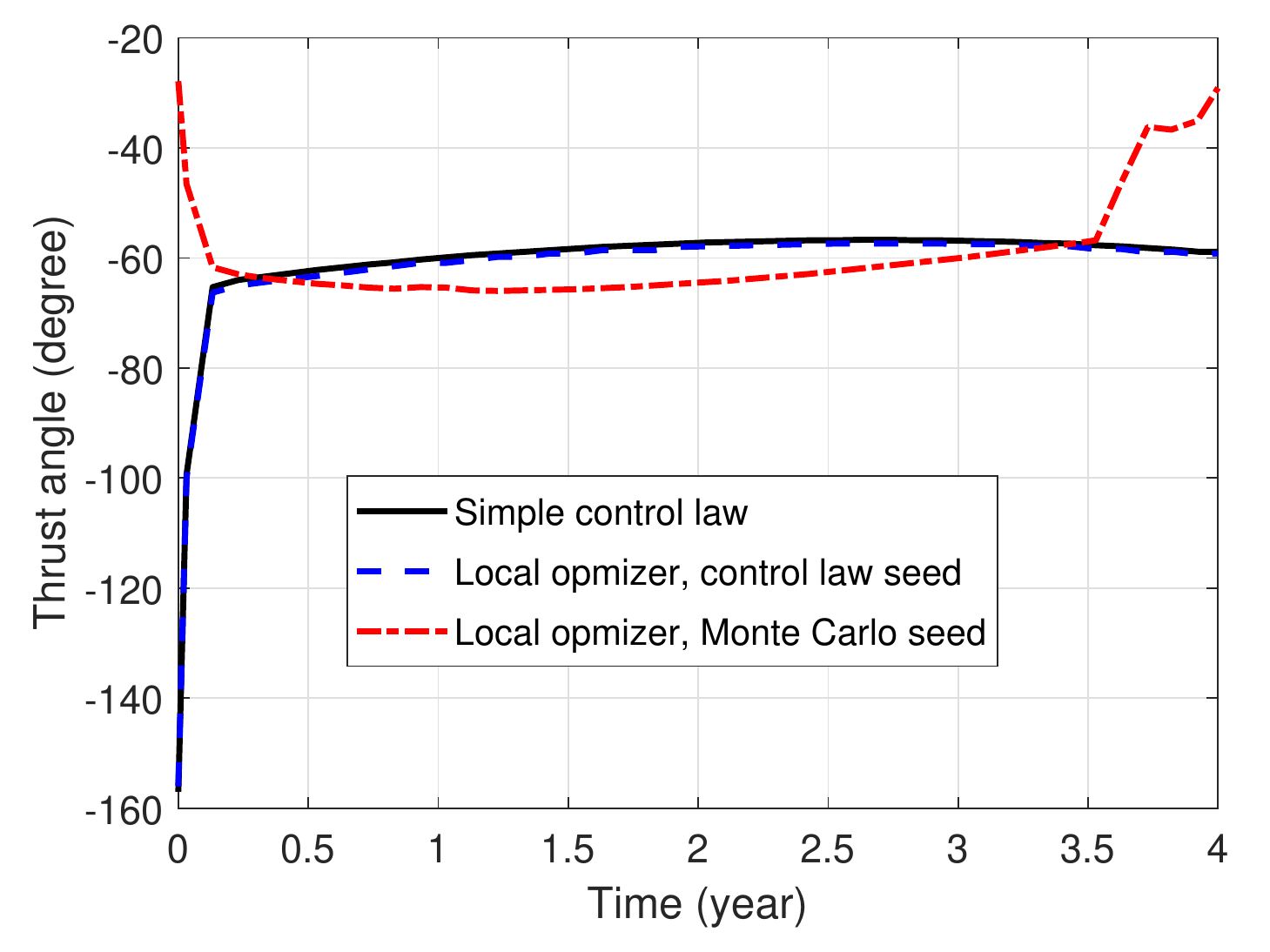}
	\caption{Thrust angle histories obtained with steering law (solid line) and optimizer (dashed lines).}
	\label{fig:global_beta}
\end{figure}

In order to assess the merits of our results, a reference solution for comparison purposes is needed. To this effect, we recomputed the Jupiter-to-Saturn LT arc with a local optimizer based on the the gradient descent technique \cite{Cauchy1847}. The algorithm uses line searches with Brent's method \cite{Brent1973} to find the smallest error along each search direction. The process is repeated for new search directions until the difference between two successive results falls below a predefined threshold. The thrust angle law is reconstructed with linear/cubic splines. The control points of the splines serve as degrees of freedom of the problem. We set the thrust arc duration to 4 years, and minimize the final excess velocity. An initial guess is needed to start the process. To determine how close our results are to a local minimum, we tried the solution from the steering law as starting configuration for the optimizer. To explore a wider solution space and assess the improvements that a global optimizer could achieve, we used a Monter Carlo approach. Multiple random seeds were generated and passed to the local optimizer. We retained 42 control points for the splines (one at the beginning of each time step of the local solution) to give the steering law and optimizer a comparable amount of flexibility. Tests showed that a better result is obtained with linear splines, as their cubic counterparts are prone to under/overshoots when the thrust angle changes rapidly. We also carried tests with fewer control points (between 5 and 10) where the cubic splines perform better. However, the reduced flexibility lead to poorer results (higher values of $V_{ex}$). Figure~\ref{fig:global_vinf} depicts the evolution of the hyperbolic excess velocity, the thrust angle is plotted in Figure~\ref{fig:global_beta}. The results correspond to the same departure conditions from Jupiter as in Figure~\ref{fig:case_879}.

The steering law achieves a relative velocity of 887 m/s after 4 years. When the local optimizer is seeded with the result of the steering law, it converges in just two iterations and returns a solution which is very similar. The final excess speed is 890 m/s. The small difference in $V_{ex}$ is due simply to the inability of the linear splines to reconstruct exactly the variation of $\beta$ from the local optimizer (the propagator uses a 13-stage RK method, so even if the values at the beginning of a time step were equal, the evolution would not be identical). For practical purposes, the two solutions are equivalent. This proves that the result of the control law is really close to one of the (many) local minimums of the solution. On the other hand, when the local optimizer is started with a random guess, the results vary wildly depending on the initial approximation and the numerical settings (e.g., step size controls and tolerances). Most attempts yield $V_{ex}$ above 1200 m/s because the algorithm gets trapped in a local extremum. With a large number of random runs we found a better solution, shown in the figures, giving a final excess velocity of 880 m/s. The optimizer is not constrained to monotonically decreasing values of the error function, as it only considers the final value. This gives it much greater flexibility in the type of trajectories it can find.

Nevertheless, the excess speed for the best solution is just 1\% below the result of the steering law. This has to be balanced against the computational cost. Computing the gradient of the objective function using a second-order centered finite-difference scheme needs 84 trajectory propagations (as there are 42 degrees of freedom). The line search requires computing about 10 additional trajectories. Then, the complete process is repeated for each search direction (usually, between 5 and 30 iterations are needed). Because the computation gets trapped in a shallow local minimum most of the time, many attempts with different initial guesses are required before a good solution is found. This means that the optimizer needs to compute tens of thousands of trajectories to match the performance of  our method, which only requires one propagation and does not rely on initial approximations (making it very robust). Obviously, computing the rates of change of the orbital elements using the planetary equations incurs an additional cost at each time step. Nevertheless, this increase in complexity does not negate the advantages of the steering law, which remains at least 3 orders of magnitude faster.

In summary, the excess speed reduction steering law offers much better computational performance and the solution is quite robust. The optimizer, on the other hand, offers the possibility of better (although not by a long margin) and more flexible solutions, but with vastly increased cost and great sensitivity to the initial settings. Therefore, both deserve a place in the toolbox of mission designers.

\section{Technological aspects and applicability}
\label{sec:app}
Since the proposed transfer strategy requires electric propulsion, the technological issues related to energy generation must be considered.
The mean solar constant drops to 50.5 W/m$^2$ at Jupiter and 15.0 W/m$^2$ at Saturn. Therefore, producing 600 W of electric power (the amount required by NEXT to impart 25 mN of thrust) using solar cells would require a very large array, which is incompatible with a small S/C mass. Radioisotope power systems (RPSs) are the only viable alternative nowadays. This technology is mature and has flown in all deep space missions to the outer solar system. The GPHS-RTG, the standard RTG used in all NASA probes, uses Pu$^{238}$ and Si-Ge thermocouples. It produces 285 W at beginning of life and 255 after 5-10 years of operation with a mass of 55 kg \cite{1987Bennett}. A suite of three GPHS-RTGs would be able to supply the thruster and provide some extra power for other systems. Moreover, the waste heat from the RTGs can be used to maintain the temperature of the S/C without draining the electric system. After arrival at Saturn, the generators can by used to drive the scientific payload.         

Many missions are being planned for the outer solar system and they all rely on RPSs \cite{VisionVoyages2011}.  
Moreover, given the shortage of Pu$^{238}$, RTGs may be replaced with more efficient advanced Stirling radioisotope generators (ASRGs) in the future. For example, Ref.~\cite{2008Oleson} proposes a mission to a Centaur body using long-lived 700 W Hall thrusters powered by six ASRGs (150 W each). This concept would deliver a 60 kg science payload with a mission duration of 10 years.

The design strategy we presented may be applied to new missions to Saturn, such as those proposed within the roadmaps of the National Research Council Decadal Survey \cite{VisionVoyages2011} and ESA Science Programme \cite{ESAScience}. As an example, NASA recently announced the approval of the Dragonfly mission to Titan \cite{2018Dragonfly} and scheduled it for launch in 2026. Then, E$^2$T is a medium-class mission to Enceladus and Titan led by ESA in collaboration with NASA and designed in response to ESA's M5 Cosmic Vision Call. E$^2$T, which will use solar-electric power, aims at a launch opportunity in 2030 \cite{2018Mitri}. Finally, TSSM, originally proposed to launch in 2020 and then superseded by the Europa Jupiter System Mission, may be revived and continues to be studied for a later launch date. All these missions could benefit from reduced excess hyperbolic velocity, which requires a much lower orbit insertion impulse than current approaches. This would enable a new, inexpensive and more flexible category of missions to Saturn.

Our excess velocity  reduction technique can be extended to missions to the outermost giant planets. For example, it may simplify substantially the gravitational capture at Neptune with an ET proposed by \cite{2018Sanmartin,2019Sanmartin}.

\section{Conclusions}
\label{sec:conclu}
This contribution presents a method to decrease substantially the hyperbolic excess speed of a spacecraft (S/C) approaching Saturn and facilitate the gravitational capture. The interplanetary trajectory includes a gravity assist at Jupiter, combined with low-thrust (LT) maneuvers. Our mission concept assumes an initial S/C mass of 1000 kg, using the performance figures of the NEXT motor, and powered by set of 3 standard radioisotope thermoelectric generators (RTGs). The RTGs can also power the scientific instruments of the probe when the electric thruster is inactive and provide waste heat to maintain the temperature of the S/C.

The thrust arc from Earth to Jupiter enables lowering the minimum launch $C_3$ from 77 to 67.25 (km/s)$^{2}$. This increases the payload injected mass by 780 kg based on the characteristics of a typical heavy launcher or, conversely, allows using a smaller booster for the same payload. The payload increase completely offsets the propellant required by the electric motor, which is just 158 kg.

The post-gravity assist (GA) thrust arc reduces the hyperbolic excess velocity at Saturn to 1 km/s using 209 kg of propellant. This low relative speed requires a Saturn insertion impulse of 148 m/s to achieve the same initial orbit as the Cassini/Huygens mission, which needed a $\Delta V$ of 622 m/s. The reduced impulse opens the door for more efficient braking methods, such as electrodynamic tethers, or even direct capture by means of a Titan flyby. The reduced excess velocity comes at the cost of a long Jupiter-to-Saturn transfer time (8 years) because a moderate eccentricity trajectory tangent to Saturn's orbit is required.

The control law used to reduce the excess velocity is, in essence, an application of the classical gradient descent technique. It is robust because it does not rely on a predefined reference path. 
The results have been compared with a gradient-based optimizer, finding that the result of steering law coincides with a local extremum of the solution. Multiple runs of the optimizer using random seeds yielded vastly different trajectories, but small differences in the minimum value of the excess velocity.
Given that the only inputs of the control law are the current and target states, it can compensate in real time for deviations from to the nominal trajectory (i.e., separate course correction maneuvers are not needed). The algorithm is very general and is not limited to the specific application described here. For example, we have used it successfully to compute optimum LT transfers between libration points of the moons of Saturn. The optimization algorithm offers a good performance on commodity CPUs. Less than 1.3 CPU seconds were required to analyze 10 000 trajectories in a current laptop processor (Core i7-8750H).

\section*{Funding sources}
The work of E. Fantino and J. Pel{\'a}ez has been supported by Khalifa University of Science and Technology's internal grants FSU-2018-07 and CIRA-2018-85. Authors of the Space Dynamics Group acknowledge also the support provided by the project entitled Dynamical Analysis of Complex Interplanetary Missions with reference ESP2017-87271-P, sponsored by Spanish Agencia Estatal de Investigaci{\'o}n (AEI) of Ministerio de Econom{\'i}a, Industria y Competitividad (MINECO) and by European Fund of Regional Development (FEDER).
The authors wish to thank the editorial board and the peer reviewers for their work and their valuable suggestions. 

\bibliography{Saturn_R1}   

\end{document}